\documentclass[epj]{svjour}
\usepackage{latexsym}
\usepackage{graphics}
\usepackage{amssymb}
\usepackage{graphicx}
\usepackage{amsmath}
\usepackage{makeidx}
\usepackage{indentfirst}
\begin{document}
\title{Nonholonomic Relativistic Diffusion and \\
Exact Solutions for Stochastic Einstein Spaces}
\author{Sergiu I. Vacaru \thanks{ sergiu.vacaru@uaic.ro, Sergiu.Vacaru@gmail.com} 
} 
%
%
\institute{
 University "Al. I. Cuza" Ia\c si, Science Department, 
 54 Lascar Catargi street, Ia\c si, Romania, 700107 
}
\date{Received: date / Revised version: date}
%
\abstract{
We develop an approach to the theory nonholonomic relativistic stochastic processes in curved spaces. 
The It\^{o} and Stratonovich calculus are formulated for spaces with conventional horizontal (holonomic) 
and vertical (nonholonomic) splitting defined by nonlinear connection structures. Geometric models of 
relativistic diffusion theory are elaborated for nonholonomic (pseudo) Riemannian manifolds and phase 
velocity spaces. Applying the anholonomic deformation method, the field equations in Einstein gravity 
and various modifications are formally integrated in general forms, with generic off--diagonal metrics 
depending on some classes of generating and integration functions. Choosing random generating 
functions we can construct various classes of stochastic Einstein manifolds. We show how  stochastic 
gravitational interactions with mixed holonomic/ nonholonomic and random variables can be modelled 
in explicit form and study their main geometric and stochastic properties. Finally, there are analyzed the 
conditions when non--random classical gravitational processes transform into stochastic ones and inversely.
\PACS{       {02.50.Ey}{ Stochastic processes}   \and
      {02.90.+p}{Other topics in mathematical methods in physics} \and 
      {02.40.-k}{Geometry, differential geometry, and topology} \and 
      {04.20.Jb}{Exact solutions}  
           } 
} 
\maketitle
\section{Introduction}

In recent years stochastic methods and the theory of (relativistic)
diffusion on Lorentz manifolds, and (for instance) Finsler/ supersymmetric
generalizations, re--attracted considerable interest and applications in
various directions of mathematics, gravity and modern cosmology and
astrophysics. During the last decade a lot of papers have been devoted to
such issues (there are many ways and approaches); for reviews and original
results in physics and mathematics, see \cite%
{chev,dunk,haba,franchi,bailleul,herrmann1,vstoch5,vstoch6}
and references therein.

In a series of our works \cite%
{vexsol2,vexsol3,vncg2}, we proved that
the gravitational field equations in various theories of gravity (general
relativity and extra dimension models, noncommutative, nonsymmetric, Finsler
etc generalizations) can be integrated in very general forms using the
so--called anholonomic deformation/ frame method. The idea of the method is
to use some distortion tensors completely defined by a given, in general,
 off--diagonal (which can not be diagonalized by coordinate transforms) metric structure
and uniquely deforming the Levi--Civita connection into ''auxiliary'' linear
connections which are also compatible and uniquely defined by the same
metric. For such a connection and with respect to some classes of nonholonomic frames of reference, the system of Einstein equations decouple split into conventional sub--systems which can be integrated ''step by step'' in very general forms. Imposing additional nonholonomic constraints on integral
varieties, we can generate exact solutions in the general relativity theory.

In explicit form, the type of exact solutions depend on the class of
generating and integration functions and considered boundary/initial
conditions, on prescribed smooth classes of coefficients and sources of
equations and on existing spacetime local and global symmetries for a chosen
topology. We have to involve additional physical arguments in order to
select a "physically important" exact solution with deterministic or random
properties. The bulk of exact solutions in gravity theories were constructed
for smooth and/or singular classes of metrics and there were analyzed
possible perturbative deformations by quasi--classical quantum contributions
and fluctuations of classical and quantum matter sources.

In this work we study new classes of stochastic gravitational processes
modelled as exact solutions of the Einstein equations (of vacuum type, with
cosmological constant and/or with various types of stochastic sources).
There are developed certain methods for the theory of stochastic calculus
and relativistic diffusion on curved spaces following the formalism of
moving frames with associated nonlinear connection structure and respective
constructions from the geometry of nonholonomic manifolds. We prove that
such new classes of stochastic solutions in gravity are derived by
integrating in general form certain systems of PDE with separation of
equations\footnote{%
it should be not confused with the method of separation of variables} and
stochastic generating functions and by imposing corresponding nonholonomic
(equivalently, anholonomic, or non--integrable) constraints on gravitational
vacuum and non--vacuum interactions.
\footnote{%
For non--experts on probability and stochastic calculus (i.e. researches
skilled in differential geometry and applications), we remember some main
concepts and ideas which are important for definition of relativistic
diffusion theory (see details in \cite{lemons,elw,ikeda,emery}): \
Physicists have abandoned determinism as a fundamental description of
reality and began developing probabilistic (that is, stochastic) models of
natural phenomena long before quantum mechanics and physics. Classical
uncertainty preceded quantum uncertainty (in our approach derived from
solutions of nonlinear systems of equations). P. Langevin, was the first to
apply Newton's second law to a ''Brownian particle'' on which the total
force included a random components. The time evolution of a random variable
is called a random or stochastic process. Thus $X(\tau )$ denotes a
stochastic process. The time $\tau $ evolution of a sure variable is called
a deterministic process and denoted $x(\tau ).$%

\textit{Random and Sure Variables: } A quantity that, under given
conditions, can assume different values is a \textit{random variable.} It
matters not whether the random variation is intrinsic and unavoidable (for
instance, a consequence of quantum effects) or an artifact of our ignorance.
A random variable is conceptually distinct from a ''certain'' or ''sure'',
i.e. deterministic, variable.

The \textit{expected value} of a random variable $X$ is a function that
turns the probabilities $P(x)$ into a sure variable called the mean of $X.$
The mean $\langle P \rangle$ parametrizes the random variable $X,$ but also do all the
\textit{moments }$\langle X^{n}\rangle $ ($n=0,1,...$) and \textit{moments
about the mean} $\langle \left( X-\langle X\rangle \right) ^{n}\rangle .$
\textit{A continuous random variable} $X(\tau )$ is completely defined by
its probability density $p(x).$%

A \textit{memoryless} process is a \textit{Markov} process. \ The Wiener
(Browninan) process, defined by the Markov propagator with a ''simple''
parameter $\rho ^{2}$ is the simplest of all Markov processes. The Wiener
process is, on its domain, everywhere continuous but nowhere smooth. This
special property makes the Wiener process dynamical equation a different
kind of mathematical object -- a \textit{stochastic differential equation}
(SDE). It is characterized by a diffusion constant. For a stochastic process
$X(t),$ the \textit{mean function} is $\mu (t)=\mathbb{E}[X(t)]$ and the
\textit{covariance function is} $k(t,t^{\prime })=\mathbb{E}[\left( X(t)-\mu
(t)\right) \left( X(t^{\prime })-\mu (t^{\prime })\right) ].$ \textit{%
Gaussian processes } are stochastic processes defined by their mean and
covariance functions.

A diffusion equation is mathematically equivalent to a stochastic dynamical
equation. In gravity theory, random variables are introduced when some
components of metric are changed by stochastic forces (generating functions)
driven by a Wiener process. Markov process can be considered but with
''some'' memory on hyperbolicity (finite speed of light), Lorentz transforms
in special relativity, and Mach principle in general relativity, as well on
Einstein equations etc. This is encoded in the definition of Stratonovich
integral on curved/nonholonomic spaces.}
 Our aim is also to formulate
special and general relativistic nonholonomic diffusion models and apply
such stochastic techniques in modern gravity theory.

With respect to exact sure and stochastic solutions of the Einstein
equation, we have to address to the general \textbf{problem of transition to
stochastic phase} of solutions for a system of nonlinear partial
differential equations (PDE) on curved manifold/ bundle spaces is the
following:

\begin{itemize}
\item \textit{If and under what conditions can an existing ''non--random''
solution (depending on certain smooth classes of generating and integration
functions) be further transformed into a stochastic solution of the same
system of equations? }

\item \textit{What types of nonholonomic constraints, boundary/inicial
conditions, and corresponding data for generating/ integration functions will
transform a nonlinear dynamical/evolution classical, or quantum process,
into a stochastic one, and inversely? }
\end{itemize}

In brief, this can be equivalently posed as a problem for formulating some
criteria for generating nonlinearly stochastic processes (for certain well
defined conditions, diffusion). In general, the solutions of systems of PDE
can be with mixed sure and random variables.

We shall construct exact stochastic solutions of Einstein equations in
explicit form for vacuum and non--vacuum gravitational configurations. Such
spacetimes are described by generic off--diagonal metrics, the nonlinear
gravitational interactions being subjected to certain types of nonholonomic
(equi\-valently, anholonomic, or non--integrable) constraints/ conditions
which may give rise, or not, to random gravitational processes. A
cosmological constant, with possible local anisotropic polarizations, and
matter sources present additional issues to be addressed.\footnote{%
A nonholonomic manifold is defined by a pair $\left( \mathbf{V},\mathcal{N}%
\right) ,$ where $\mathbf{V}$ is a manifold and $\mathcal{N}$ \ is a
non--integrable distribution. We have to involve in our research certain
methods from the geometry of nonholonomic distributions with associated
nonlinear connection structure (geometry of nonholonomic manifolds) and
introduce on Minkowski and (pseudo) Riemannian spaces some types of
Lagrange--Finsler parametrizations of metrics and connections because only
using such variables we can construct general exact off--diagonal solutions
in gravity and formulate an unified mathematical approach to classical field
interactions and random processes with mixed holonomic/ nonholonomic /
stochastic components.}

A partner work \cite{vstoch2e} contains developments of the results of this
paper for stochastic Ricci--Finsler flows and gravitational interactions/
evolutions modeling porous media and self--organized criticality.

\subsection{Related directions}

In order to avoid possible ambiguities with the formalism and terminology
used in different approaches, we mention here some alternative methods with
stochastic/ diffusion in curved spaces and random variables in gravity:

\begin{enumerate}
\item The theory of stochastic semiclassical gravity (in brief, \textit{%
stochastic gravity}) is based on the Einstein--Langeven equation with
additional sources due to the noise kernel. It was naturally constructed
from semiclassical gravity and quantum field theory in curved spacetimes and
non--equilibrium statistical mechanics, see a comprehensive review \cite{hu}
and references therein. For short, stochastic gravity includes also its
fluctuations, for instance, of quantum fields in curved spaces and explore
how the metric fluctuations are induced and seed the structure of the
universe, how such processes affect the back reaction of Hawking radiance in
black hole dynamics and the black hole horizons, trans--Planckian physics
etc.

\item The theory of stochastic (diffusion) equations on curved manifolds
with local Euclidian signature\footnote{%
in this paper we shall use equivalently the terms "space" and/or
"spacetimes" for certain (pseudo/semi) Riemannian geometries, endowed with
metrics of necessary signature, nonholonomic distributions with associated
anholonomic frames, and (auxiliary) generalized connections} is a well
developed topic (see e.g. \cite{elw,ikeda,emery}), see also
related applications to locally anisotropic kinetic processes, gravity and
astrophysics \cite{vstoch5,vstoch6} and an alternative approach to
anisotropic diffusion \cite{christensen}. For such constructions, random
processes were considered on classical (manifolds) inducing possible matter
field sources for gravitational filed equations.
\end{enumerate}

Here we also note that various stochastic methods were also applied and
developed for quantizing gauge and gravitational fields \cite{sq1,sq2,sq3},
see also a recent attempt for stochastic quantization of Ho\v{r}%
ava--Lifshitz gravity \cite{sq4}. In this work, we shall restrict our
considerations only to nonlinearly defined classical (non--quantum)
gravitational processes generating exact solutions of the Einstein (not
obligatory in the form of Einstein--Langeven) equations.

\subsection{Goals of the paper:}
Conventionally, there are three main purposes:
\begin{enumerate}
\item To provide an introduction into the theory of stochastic processes on
nonolonomic Euclidian and Riemannian manifolds.

\item To generalize the theory of stochastic equations in relativistic form
for nonholonomic (pseudo) Riemannian manifolds in a form allowing to study
mixed sure (non--random) and stochastic gravitational processes.

\item To prove that prescribing stochastic generating and integration
functions for ''formal'' general solutions of Einstein equations we define
three types (so--called, horizontal, vertical and horizontal--vertical ones)
of stochastic (nonholonomic) spacetimes. It will be shown how to construct
general stochastic gravitational solutions with Killing symmetries and
extensions to non--Killing ones and formulated the criteria when the
non--stochastic solutions may transform into some stochastic ones, and
inversely.
\end{enumerate}

We organize the exposition as follows:  In section \ref{ssenrm} we summarize the It\^{o} and Stratonovich stochastic
calculus and diffusion theory on nonholonomic Euclidian and Riemannian
manifolds. We use a synthesis of geometric and stochastic methods originally elaborated for the diffusion theory on nonholonomic vector bundles spaces and in Lagrange--Finsler geometry.
  Section \ref{snrdife} is devoted to the theory of nonholonomic special and general relativistic diffusion. A formalism of adapting the constructions to the nonlinear connection structure play a key role in distinguishing
nonlinear gravitational and random gravitational processes and keeping
certain similarity with former constructions for phase velocity spaces and
moving frame method.  In section \ref{seses}, we generalize the anholonomic deformation method of constructing exact solutions in gravity in order to include into the scheme
the possibility to generate stochastic metrics and nonlinear and linear
connections. We also analyze explicit conditions/criteria when such
stochastic gravitational fields are of vacuum type, with gravitational
polarizations and/or cosmological constants and state the possibility to
extend the constructions for nontrivial matter sources.
 A summary of results and concluding remarks are given in section \ref{sconcl}%
.
\section{Stochastic Processes and Nonholonomic Mani\-folds}
\label{ssenrm}

In this section, we provide an introduction into the theory of stochastic
differential equations on nonholonomic (pseudo) Riemannian manifolds \cite%
{vexsol2}. We assume that the reader is familiar with the
concepts and basic results on stochastic calculus, Brwonian motion and
stochastic processes ''rolled'' on curved (usually, Riemannian) spaces \cite%
{elw,ikeda,emery}.

\subsection{Geometric preliminaries}

We consider a 4--d manifold (space, or spacetime) $\mathbf{V}$ endowed with
a metric
\begin{equation}
\mathbf{g}=g_{\alpha \beta }(u^{\gamma })du^{\alpha }\otimes du^{\beta }
\label{metr}
\end{equation}%
of necessary signature $(\pm ,\pm ,\pm ,\pm )$ \ when local coordinates are
parametrized in the form $u^{\alpha }=(x^{i},y^{a}),$ where $%
x^{i}=(x^{1},x^{2})$ and $y^{a}=\left( y^{3}=v,y^{4}=y\right) .$\footnote{%
In this paper, we shall consider metrics with local signature $(+,+,-,+)$
and/or $(+,+,+,-)$ in order to follow conventions from our previous results
on exact solutions with local anisotropy \cite%
{vexsol2,vexsol3}.} Indices $i,j,k,...=1,2$ and $%
a,b,c,...=3,4$ are for a conventional $(2+2)$--splitting of dimension when
the general (small Greek) abstract/coordinate indices when $\alpha ,\beta
,\ldots $ run values $1,2,3,4.$

For our purposes, we split a metric (\ref{metr}) in the form
\begin{equation}
\ \mathbf{g}\mathbf{=}g_{ij}{dx^{i}\otimes dx^{j}}%
+h_{ab}(dy^{a}+N_{k}^{a}dx^{k}){\otimes }(dy^{b}+N_{k}^{b}dx^{k}),
\label{dm}
\end{equation}%
and parametrize as
\begin{eqnarray}
\ ^{\eta }\mathbf{g} &\mathbf{=}&\eta _{i}(x^{k},v)\ ^{\circ }g_{i}(x^{k},v){%
dx^{i}\otimes dx^{i}}+\eta _{a}(x^{k},v)\ ^{\circ }h_{a}(x^{k},v)\mathbf{e}%
^{a}{\otimes }\mathbf{e}^{a},  \label{gsol} \\
\mathbf{e}^{3} &=&dv+\eta _{i}^{3}(x^{k},v)\ ^{\circ }w_{i}(x^{k},v)dx^{i},\
\mathbf{e}^{4}=dy^{4}+\eta _{i}^{4}(x^{k},v)\ ^{\circ }n_{i}(x^{k},v)dx^{i}.
\notag
\end{eqnarray}%
In (\ref{gsol}), we consider that $g_{ij}=diag[g_{i}=\eta _{i}\ ^{\circ
}g_{i}]$ and $h_{ab}=diag[h_{a}=\eta _{a}\ ^{\circ }h_{a}]$ and $%
N_{k}^{3}=w_{i}=\eta _{i}^{3}\ ^{\circ }w_{i}$ and $N_{k}^{4}=n_{i}=\eta
_{i}^{4}\ ^{\circ }n_{i}.$ The gravitational 'polarizations' $\eta _{\alpha
} $ and $\eta _{i}^{a}$ determine nonholonomic deformations of metrics, $\
^{\circ }\mathbf{g}\mathbf{=}[\ ^{\circ }g_{i},\ ^{\circ }h_{a},\ ^{\circ
}N_{k}^{a}]\rightarrow \ ^{\eta }\mathbf{g}\mathbf{=}[\
g_{i},h_{a},N_{k}^{a}]$ and can be defined by functions of necessary smooth
class and/or any random (stochastic) variables.\footnote{%
Such transforms, with deformations of the frame, metric, connections and
other fundamental geometric structures, are more general than those
considered for the Cartan's moving frame method when the geometric objects
are re--defined equivalently with respect to various classes of systems of
reference.}

We use a boldface symbol $\mathbf{V}$ for nonholonomic manifolds
(equivalently, spaces) when a Whitney splitting of tangent space $T\mathbf{V}
$ is defined,%
\begin{equation}
T\mathbf{V}=h\mathbf{V\oplus }v\mathbf{V,}  \label{whitney}
\end{equation}%
with conventional horizontal (h) and vertical (v) subspaces, respectively, $h%
\mathbf{V}$ and $v\mathbf{V.}$ Such a geometric object defines a nonlinear
connection (N--connection) structure $\mathbf{N,}$ stated locally by a set
of coefficients $\{N_{k}^{a}\}$ with respect to a corresponding coordinate
basis. It should be emphasized here that we can introduce on (pseudo)
Riemannian manifolds various types of nonholonomic distributions and
N--connections, for instance, with $2+2$ splitting, following the principle
of covariance allowing us to consider any equivalently any system of
reference/coordinates.

A N--connection structure $\mathbf{N}=\{N_{k}^{a}\}$ is present in (\ref{dm}%
) via $N$--adapted frame, $\mathbf{e}_{\alpha },$ and dual frame, $\mathbf{e}%
_{\ }^{\beta },$ structures (i.e. N--elongated partial derivatives,
respectively, differentials)
\begin{eqnarray}
\mathbf{e}_{\alpha } &\doteqdot &\left( \mathbf{e}_{i}=\partial
_{i}-N_{i}^{a}\partial _{a},e_{b}=\partial _{b}=\frac{\partial }{\partial
y^{b}}\right) ,  \label{ddr} \\
\mathbf{e}_{\ }^{\beta } &\doteqdot &\left( e^{i}=dx^{i},\mathbf{e}%
^{a}=dy^{a}+N_{i}^{a}dx^{i}\right) .  \label{ddf}
\end{eqnarray}%
Such local bases satisfy nonholonomic relations of type
\begin{equation*}
\left[ \mathbf{e}_{\alpha },\mathbf{e}_{\beta }\right] =\mathbf{e}_{\alpha }%
\mathbf{e}_{\beta }-\mathbf{e}_{\beta }\mathbf{e}_{\alpha }=\mathbf{w}_{\
\alpha \beta }^{\gamma }\left( u\right) \mathbf{e}_{\gamma },
\end{equation*}%
with nontrivial anholonomy coefficients $\mathbf{w}_{\beta \gamma }^{\alpha
}\left( u\right) ,$
\begin{equation*}
\mathbf{w}_{~ji}^{a}=-\mathbf{w}_{~ij}^{a}=\Omega _{ij}^{a}=\mathbf{e}%
_{j}N_{i}^{a}-\mathbf{e}_{i}N_{j}^{a},\ \mathbf{w}_{~ia}^{b}=-\mathbf{w}%
_{~ai}^{b}=\partial _{a}N_{i}^{b}
\end{equation*}%
which is also related to the concept of nonholonomic manifold \cite%
{vexsol2,vexsol3}, i.e. a manifold endowed with a
non--integrable distribution (in particular, with N--connection structure).

On a spacetime $\mathbf{V,}$ there is an infinite number of metric
compatible linear connections $D,$ satisfying the conditions $D\mathbf{g}=0,$
and completely defined by a metric $\mathbf{g}$ (\ref{dm}). A subclass of
such linear connections can be adapted to a chosen N--connection structure $%
\mathbf{N,}$ when the splitting (\ref{whitney}) is preserved under
parallelism, and called distinguished connections (in brief, d--connections).%
\footnote{%
On spaces endowed with N--connection structure, there are used the terms
distinguished vectors/ forms / tensors etc (in brief, d--vectors, d--forms,
d--tensors etc) in order to emphasize that the geometric constructions are
adapted to the N--connection structure, i.e. preserving h-- and v--splitting.%
} A general d--connection is denoted $\mathbf{D}=(hD,vD),$ being
distinguished into, respectively, h- and v--covariant derivatives, $hD$ and $%
vD.$ We proved \cite{vexsol2,vexsol3} that the Einstein equations in various gravity theories decouple
and became integrable in very general forms for the so--called canonical d--connection, $\widehat{\mathbf{D}}%
. $ With respect to N--adapted bases (\ref{ddr}) and (\ref{ddf}), $\widehat{%
\mathbf{D}}$ is computed to have the coefficients $\widehat{\mathbf{\Gamma }}%
_{\ \alpha \beta }^{\gamma }=(\widehat{L}_{jk}^{i},\widehat{L}_{bk}^{a},%
\widehat{C}_{jc}^{i},\widehat{C}_{bc}^{a}),$ for $h\widehat{D}=\{\widehat{L}%
_{jk}^{i},\widehat{L}_{bk}^{a}\}$ and $v\widehat{D}=\{\widehat{C}_{jc}^{i},%
\widehat{C}_{bc}^{a}\},$ with
\begin{eqnarray}
\widehat{L}_{jk}^{i} &=&\frac{1}{2}g^{ir}\left(
e_{k}g_{jr}+e_{j}g_{kr}-e_{r}g_{jk}\right),\ \widehat{L}_{bk}^{a} =e_{b}(N_{k}^{a})+\frac{1}{2}h^{ac}\left(
e_{k}h_{bc}-h_{dc}\ e_{b}N_{k}^{d}-h_{db}\ e_{c}N_{k}^{d}\right) ,  \label{candcon}   \\
\widehat{C}_{jc}^{i} &=&\frac{1}{2}g^{ik}e_{c}g_{jk},\ \widehat{C}_{bc}^{a}=%
\frac{1}{2}h^{ad}\left( e_{c}h_{bd}+e_{c}h_{cd}-e_{d}h_{bc}\right) .  \notag
\end{eqnarray}%
The d--connection $\widehat{\mathbf{D}}$ and its torsion $\widehat{\mathcal{T%
}}=\{\widehat{\mathbf{T}}_{\ \alpha \beta }^{\gamma }\equiv \widehat{\mathbf{%
\Gamma }}_{\ \alpha \beta }^{\gamma }-\widehat{\mathbf{\Gamma }}_{\ \beta
\alpha }^{\gamma }; 
\widehat{T}_{\ jk}^{i},\widehat{T}_{\ ja}^{i},\widehat{T}_{\ ji}^{a},%
\widehat{T}_{\ bi}^{a},\widehat{T}_{\ bc}^{a}\},$ where the nontrivial
coefficients
\begin{equation}
\widehat{T}_{\ jk}^{i} =\widehat{L}_{jk}^{i}-\widehat{L}_{kj}^{i},\widehat{%
T}_{\ ja}^{i}=\widehat{C}_{jb}^{i},\widehat{T}_{\ ji}^{a}=-\Omega _{\
ji}^{a},\  \widehat{T}_{aj}^{c} =\widehat{L}_{aj}^{c}-e_{a}(N_{j}^{c}),\widehat{T}_{\
bc}^{a}=\ \widehat{C}_{bc}^{a}-\ \widehat{C}_{cb}^{a},   \label{dtors} 
\end{equation}%
are completely defined by the coefficients of metric $\mathbf{g}$ (\ref{dm})
following the conditions that $\widehat{\mathbf{D}}\mathbf{g=}0$ and the
''pure'' horizontal and vertical torsion coefficients are zero, i. e. $%
\widehat{T}_{\ jk}^{i}=0$ and $\widehat{T}_{\ bc}^{a}=0.$

Any geometric construction for the canonical d--connection $\widehat{\mathbf{%
D}}$ can be re--defined equivalently into a similar one with the
Levi--Civita connection following formula
\begin{equation}
\Gamma _{\ \alpha \beta }^{\gamma }=\widehat{\mathbf{\Gamma }}_{\ \alpha
\beta }^{\gamma }+Z_{\ \alpha \beta }^{\gamma },  \label{deflc}
\end{equation}%
where the distortion tensor $Z_{\ \alpha \beta }^{\gamma }$ is constructed
in a unique form\footnote{%
we can see this from explicit formulas
\begin{eqnarray*}
\ Z_{jk}^{a} &=&-\widehat{C}_{jb}^{i}g_{ik}h^{ab}-\frac{1}{2}\Omega
_{jk}^{a},~Z_{bk}^{i}=\frac{1}{2}\Omega _{jk}^{c}h_{cb}g^{ji}-\Xi _{jk}^{ih}~%
\widehat{C}_{hb}^{j},\ 
Z_{bk}^{a} =\ ~^{+}\Xi _{cd}^{ab}~\widehat{T}_{kb}^{c},\ Z_{kb}^{i}=\frac{1}{%
2}\Omega _{jk}^{a}h_{cb}g^{ji}+\Xi _{jk}^{ih}~\widehat{C}_{hb}^{j},\
Z_{jk}^{i}=0, \\
\ Z_{jb}^{a} &=&-~^{-}\Xi _{cb}^{ad}~\widehat{T}_{jd}^{c},\ Z_{bc}^{a}=0,\
Z_{ab}^{i}=-\frac{g^{ij}}{2}\left[ \widehat{T}_{ja}^{c}h_{cb}+\widehat{T}%
_{jb}^{c}h_{ca}\right] ,
\end{eqnarray*}%
for $\ \Xi _{jk}^{ih}=\frac{1}{2}(\delta _{j}^{i}\delta
_{k}^{h}-g_{jk}g^{ih})$ and $~^{\pm }\Xi _{cd}^{ab}=\frac{1}{2}(\delta
_{c}^{a}\delta _{d}^{b}+h_{cd}h^{ab})$} from the coefficients of a metric $%
\mathbf{g}_{\alpha \beta }.$

\subsection{$h$-- and $v$--adapted Euclidian diffusion}

A nonholonomic manifold with conventional h- and v--splitting is with local
fibered structure and similar to a vector/tangent bundle enabled with
N--connection structure.  In this work, we develop the approach in relativistic form (for
holonomic spaces with local pseudo--Euclidean signature when the curved spacetime $%
\mathbf{V}$ has a tangent space with splitting of dimension $n+m,$ for $%
n,m\geq 2$.

A distinguished Wiener process (in brief, Wiener d--process) of dimension $%
n+m$ is defined locally by a couple of elementary (Wiener) h-- and
v--processes $\mathcal{W}^{\alpha }(\tau )=\left( \mathcal{W}^{i}(\tau ),%
\mathcal{W}^{a}(\tau )\right) ,$ where $\tau $ (in particular, we can take $%
\tau =t$ to be a time like \ parameter).\footnote{%
\ In Euclidian space, the coefficients of a $\left( n+m\right) $%
--dimensional such Wiener process $d\mathcal{W}^{\alpha }=\mathcal{W}%
^{\alpha }(\tau +\bigtriangleup \tau )-\mathcal{W}^{\alpha }(\tau )$ are
defined for the probability density $\mathcal{P}(\mathcal{W}^{\alpha })=%
\frac{1}{\sqrt{2\rho \pi \bigtriangleup \tau }}\exp \left( -\frac{\left[
\mathcal{W}^{\alpha }(\tau )\right] ^{2}}{2\rho \bigtriangleup \tau }\right)
$ when the expectations $\langle \mathcal{W}^{\alpha }\rangle =0$ and $%
\langle \mathcal{W}^{\alpha }(\ ^{1}\tau )\mathcal{W}^{\beta }(\ ^{1}\tau +\
^{2}\tau )\rangle =\rho (\ ^{2}\tau )\ \delta ^{\alpha \beta },$ for $\delta
^{\alpha \beta }$ being the Kronecker symbol.} We consider a random
(stochastic) curve on $\mathbf{V}$ lifted to the horizontal curve on the
frame of orthonormalized bundles $O(\mathbf{V})$ related by frame
transforms, $\mathbf{e}_{\alpha ^{\prime }}=\mathbf{e}_{\ \alpha ^{\prime
}}^{\underline{\alpha }}(u)\partial _{\underline{\alpha }}=\mathbf{e}_{\
\alpha ^{\prime }}^{\alpha }(u)\mathbf{e}_{\alpha },$ to N--adapted bases $%
\mathbf{e}_{\alpha }$ (\ref{ddr}) when by $\partial _{\underline{\alpha }%
}=\partial /\partial u^{\underline{\alpha }}=(\partial _{\underline{i}%
}=\partial /\partial x^{\underline{i}},\partial _{\underline{a}}=\partial
/\partial y^{\underline{a}})$ we denote a local coordinate base (if it will
be necessary, we shall underline indices for coordinate bases; primed
indices will be used for coordinates with respect to orthonormalized frames
of reference).

\subsubsection{It\^{o} d--calculus}

A diffusion distinguished process (d--process) in an Euclidian space is
described \ by a couple of horizontal and vertical stochastic differential
equations,%
\begin{equation}
d\mathbf{U}^{\alpha }=\sigma _{\alpha ^{\prime }}^{\alpha }(\tau ,\mathbf{U}%
)\delta \mathcal{W}^{\alpha ^{\prime }}+b^{\alpha }(\tau ,\mathbf{U})d\tau
\label{stdeq}
\end{equation}%
where $\mathbf{U}=\left( hU,vU\right) \in \mathbb{R}^{n+m}$ is a stochastic
d--process with $\mathbf{U}(0)=\mathbf{u},$ for $\mathbf{u}=\{u^{\beta
}=(x^{j},y^{c})\},$ with parameter (time like variable, or temperature, $%
\tau \geq 0$). The given values $\sigma _{\alpha ^{\prime }}^{\alpha }$ and $%
b^{\alpha }$ are called respectively \ the diffusion coefficients and the
drift coefficients. \ It is possible to transform (\ref{stdeq}) into an
integral equation $
\mathbf{U}_{\tau }^{\alpha }=\mathbf{U}_{0}^{\alpha }+\int\limits_{0}^{\tau
}\sigma _{\alpha ^{\prime }}^{\alpha }(\varsigma ,\mathbf{U})\delta \mathcal{%
W}_{\varsigma }^{\alpha ^{\prime }}+\int\limits_{0}^{\tau }b^{\alpha
}(\varsigma ,\mathbf{U})d\varsigma$,
 i.e. the It\^{o} stochastic integral adapted to h- and v--splitting,
defining an It\^{o} process as a Markovian process (see details in \cite%
{elw,ikeda,emery,herrmann1}). In the above formulas we write $%
\delta \mathcal{W}_{\varsigma }^{\alpha ^{\prime }}$ instead of $d\mathcal{W}%
_{\varsigma }^{\alpha ^{\prime }}$ in order to emphasize that the approach
is with N--elongated partial derivatives and differentials, (\ref{ddr}) and (%
\ref{ddf}), instead of usual ones.

If $\mathbf{U}_{\tau }^{\alpha }$ is an It\^{o} process, then a function $%
\mathbf{Y}_{\tau }^{\alpha }=f\left( \mathbf{U}_{\tau }^{\alpha }\right) $
is also an It\^{o} process when the It\^{o} N--adapted formula for
stochastic differential $\delta f=\mathbf{A}f$ with associated diffusion
d--operator $\mathbf{A}=hA\oplus vA,$
\begin{equation}
\mathbf{A} =\frac{\rho }{2}\sum\limits_{\alpha ^{\prime }=1}^{n+m}\{\frac{1%
}{2}\sigma _{\alpha ^{\prime }}^{i}(\tau ,\mathbf{U})\sigma _{\alpha
^{\prime }}^{j}(\tau ,\mathbf{U})\left( \mathbf{e}_{i}\mathbf{e}_{j}+\mathbf{%
e}_{j}\mathbf{e}_{i}\right) f +\sigma _{\alpha ^{\prime }}^{a}(\tau ,\mathbf{U})\sigma _{\alpha ^{\prime
}}^{b}(\tau ,\mathbf{U})e_{a}e_{b}f+b^{\alpha }(\tau ,\mathbf{U})\mathbf{e}%
_{\alpha }f\},  \label{aopito}
\end{equation}%
where, for instance, 
 $\ hA=\frac{\rho }{2}\sum\limits_{\alpha ^{\prime }=1}^{n}\{\frac{1}{2}\sigma
_{\alpha ^{\prime }}^{i}(\tau ,\mathbf{U})\sigma _{\alpha ^{\prime
}}^{j}(\tau ,\mathbf{U})(\mathbf{e}_{i}\mathbf{e}_{j}+\mathbf{e}_{j}\mathbf{e%
}_{i})f+b^{i}(\tau ,\mathbf{U})\mathbf{e}_{i}f\}$. 
 Such an operator contains the second derivative and the It\^{o} stochastic
d--differential $\delta f$ \ does not satisfy usual properties for linear
operators which makes the theory more sophisticate.

For any stochastic N--adapted process $\mathbf{U}_{\tau }^{\alpha }$ $\in
\mathbb{R}^{n+m},$ we can introduce the probability density function $\phi
(\tau ,\mathbf{u})$ and determine how it evolves with time/temperature
parameter $\tau .$ This function allows us to compute the probability $\Pr $
that a realization of a set of variables $U_{1},...U_{n+m}$ falls inside a
domain $\mathcal{U}$ in the $(n+m)$--dimensional space of such variables,%
\begin{equation}
\Pr \left( \mathbf{U}_{\tau }^{\alpha }\in \mathcal{U}\right) :=\int\limits_{%
\mathcal{U}}\phi (\tau ,\mathbf{u})dx^{1}...dx^{n}\delta y^{n+1}...\delta
y^{n+m}.  \label{pr}
\end{equation}%
We can calculate the expected value $\ ^{u}E[f(\mathbf{U}_{\tau })]$ for any
function\footnote{%
in physical applications, we can consider a smooth class or any class for
which a corresponding integration procedure is defined} $f$ \ of $\mathbf{U,}
$%
\begin{equation}
\ ^{u}E[f(\mathbf{U}_{\tau })]:=\int\limits_{\mathcal{U}}f(\mathbf{u})\phi
(\tau ,\mathbf{u})\delta u^{1}...\delta u^{n+m}.  \label{mexp}
\end{equation}

If we define $\underline{f}(\tau ,\mathbf{u}):=\ ^{u}E[f(\mathbf{U}_{\tau
})],$ such a function is subjected to the condition (in literature, it is
called the Focker--Plank, or the forward Kolmogorov, equation)%
\begin{eqnarray}
\partial _{\tau }\underline{f}(\tau ,\mathbf{u}) &=&\mathbf{A}\underline{f}%
(\tau ,\mathbf{u}),  \label{fpeqito} \\
\underline{f}(0,\mathbf{u}) &=&f(\tau ,\mathbf{u}),  \notag
\end{eqnarray}%
where $\partial _{\tau }=\partial /\partial \tau $ and the d--operator $%
\mathbf{A}$ is defined by (\ref{aopito}). Physical applications are usually
considered following exact/approximate solutions of such equations.\footnote{%
\label{fn1}Similar constructions are possible for the (Hermitian adjoint)
d--operator \thinspace $\ ^{\ast }\mathbf{A}$ of $\mathbf{A}$ (the formulas
can be proven using (\ref{pr}) and (\ref{mexp})), $\partial _{\tau }\phi
(\tau ,\mathbf{u})=\,\ ^{\ast }\mathbf{A}\phi (\tau ,\mathbf{u});$ for
arbitrary function $f(\mathbf{u}),$ $\int\limits_{\mathcal{U}}f(\mathbf{u})\
^{\ast }\mathbf{A}\phi (\tau ,\mathbf{u})du^{1}...du^{n+m}=\int\limits_{%
\mathcal{U}}f(\mathbf{u})\partial _{\tau }\phi (\tau ,\mathbf{u}%
)du^{1}...du^{n+m},$ for $\ ^{\ast }\mathbf{A}=\frac{\rho }{2}%
\sum\limits_{\alpha ^{\prime }=1}^{n+m}\{\frac{1}{2}\sigma _{\alpha ^{\prime
}}^{i}(\tau ,\mathbf{U})\sigma _{\alpha ^{\prime }}^{j}(\tau ,\mathbf{U}%
)\left( \mathbf{e}_{i}\mathbf{e}_{j}+\mathbf{e}_{j}\mathbf{e}_{i}\right)
f+\sigma _{\alpha ^{\prime }}^{a}(\tau ,\mathbf{U})\sigma _{\alpha ^{\prime
}}^{b}(\tau ,\mathbf{U})e_{a}e_{b}f-\mathbf{e}_{\alpha }b^{\alpha }(\tau ,%
\mathbf{U})f\}.$}

Finally, we provide the formula $\mathbf{A}f=\lim_{\tau \rightarrow 0}\frac{%
\ ^{u}E[f(\mathbf{U}_{\tau })]-f(\mathbf{u})}{\tau },$ defined for any
suitable function $f,$ where $\mathbf{u}=\mathbf{U}_{\tau =0} $ is the
initial point of N--adapted stochastic process $\mathbf{U}_{\tau }.$ Under
the conditions that above integral formulas hold true, we can say that a
diffusion generator/d--operator $\mathbf{A,}$ and its adjoint $\ ^{\ast }%
\mathbf{A,}$ of $\mathbf{U}_{\tau }$ is associated to an It\^{o} d--process.

\subsubsection{Stratonovich d--calculus}

There is an equivalent \ reformulation of the stochastic calculus by using
the Stratonovich integral\footnote{%
even such integrals do not result in Markovian processes}, which is more
convenient for various curved spacetime generalizations. In N--adapted form,
we write
\begin{equation*}
d\mathbf{U}^{\alpha }=\widetilde{\sigma }_{\alpha ^{\prime }}^{\alpha }(\tau
,\mathbf{U})\circ \delta \mathcal{W}^{\alpha ^{\prime }}+\widetilde{b}%
^{\alpha }(\tau ,\mathbf{U})d\tau ,
\end{equation*}%
where we put ''$\sim "$ and ''$\circ "$ in order to not confuse this
interpretation with that given by (\ref{stdeq}) for the It\^{o} approach. To
get equivalent formulations of stochastic calculus is possible if we identify%
\begin{equation*}
\widetilde{\sigma }_{\alpha ^{\prime }}^{\alpha }(\tau ,\mathbf{U})=\sigma
_{\alpha ^{\prime }}^{\alpha }(\tau ,\mathbf{U}),\ \widetilde{b}^{\alpha
}(\tau ,\mathbf{U})=b^{\alpha }(\tau ,\mathbf{U})-\frac{\rho }{2}%
\sum\limits_{\alpha ^{\prime }=1}^{n+m}\sigma _{\alpha ^{\prime }}^{\beta
}(\tau ,\mathbf{U})\mathbf{e}_{\beta }\sigma _{\alpha ^{\prime }}^{\alpha
}(\tau ,\mathbf{U}).
\end{equation*}%
Such a re--definition of drift coefficients results in a linear (on partial
derivatives) operator for the associated diffusion/generator d--operator
(compare, for instance, with (\ref{aopito}) where there are contained second
partial derivatives), \ for the Stratonovich interpretation \ written in the
\ form
\begin{equation}
\widetilde{\mathbf{A}}=\frac{\rho }{2}\sum\limits_{\alpha ^{\prime }=1}^{n+m}%
\mathbf{L}_{\alpha ^{\prime }}\mathbf{L}_{\alpha ^{\prime }}+\mathbf{L}_{0},
\label{astrat}
\end{equation}%
where $\mathbf{L}_{\alpha ^{\prime }}=\sigma _{\alpha ^{\prime }}^{\beta
}(\tau ,\mathbf{u})\mathbf{e}_{\beta }$ and $\mathbf{L}_{0}=\widetilde{b}%
^{\alpha }(\tau ,\mathbf{u})\mathbf{e}_{\beta }$ are called the fundamental
(for a diffusion d--process) d--vector fields.

The associated N--adapted Focker--Plank equation (compare with (\ref{fpeqito}%
)) in the Stratonovich approach is
\begin{equation*}
\partial _{\tau }\phi (\tau ,\mathbf{u})=\frac{\rho }{2}\sum\limits_{\alpha
^{\prime }=1}^{n+m}\mathbf{e}_{\beta }\sigma _{\alpha ^{\prime }}^{\beta
}(\tau ,\mathbf{u})\mathbf{e}_{\gamma }\left[ \sigma _{\alpha ^{\prime
}}^{\gamma }(\tau ,\mathbf{u})\phi (\tau ,\mathbf{u})\right] -\mathbf{e}%
_{\alpha }\widetilde{b}^{\alpha }(\tau ,\mathbf{u})\phi (\tau ,\mathbf{u}).
\end{equation*}

\subsection{Diffusion on nonholonomic manifolds}

\label{ssdnhm}The theory of stochastic differential equations on a
nonholonomic manifold $\mathbf{V}$ enabled with a metric compatible
d--connection can be constructed similarly to the case of $n+m$ dimensional
Riemannian spaces if we work with N--adapted frames and co--frames (\ref{ddr}%
) and (\ref{ddf}). We use N--adapted variables/coordinates $\mathbf{r}=(%
\mathbf{u,e})=(u^{\alpha },\mathbf{e}_{\ \beta }^{\beta ^{\prime }})\in O(%
\mathbf{V})$ and their frame/coordinate transforms. The infinitesimal motion
of a smooth curve $u^{\alpha }(\tau )\in \mathbf{V}$ is lifted naturally to
that of $\ \gamma ^{\alpha ^{\prime }}(\tau )\in O(\mathbf{V})$ using the
ordinary differential equations for N--adapted to (\ref{whitney}) parallel
transport,%
\begin{equation*}
\delta u^{\alpha }=\mathbf{e}_{\ \alpha ^{\prime }}^{\alpha }(u^{\beta
})\delta \gamma ^{\alpha ^{\prime }}\mbox{ and }\delta \mathbf{e}_{\ \alpha
^{\prime }}^{\alpha }(u^{\mu })=-\widehat{\mathbf{\Gamma }}_{\ \beta \nu
}^{\alpha }(u^{\mu })\mathbf{e}_{\ \alpha ^{\prime }}^{\nu }(u^{\mu })\delta
u^{\beta },
\end{equation*}%
where we use symbols $\delta u^{\alpha },\delta \mathbf{e}_{\ \alpha
^{\prime }}^{\alpha }$ etc instead of respective $du^{\alpha },d\mathbf{e}%
_{\ \alpha ^{\prime }}^{\alpha }...,$ in order to emphasize that in our
constructions we use N--elongated partial derivatives and differentials. The
coefficients of the canonical d--connection $\widehat{\mathbf{\Gamma }}_{\
\beta \nu }^{\alpha }$ are computed following formulas (\ref{candcon}).

A curve $\mathbf{r}(\tau )=(\mathbf{u}(\tau )\mathbf{,e}(\tau ))$ is
considered as the horizontal lift (it should be not confused with the \
h--splitting considered in previous sections) of a curve $\mathbf{u}(\tau )$
to the nonholonomic bundle $O(\mathbf{V})$ (modelled by atlases of carts
covered with \ Euclidean spaces $\mathbb{R}^{(n+m)^{2}+n+m}$ and a
''horizontal'' curve $\gamma ^{\alpha ^{\prime }}(\tau )$ in the tangent
spaces, which can be identified locally with an Euclidean spaces $\mathbb{R}%
^{n+m}).$ \ Using such lifts, we can define stochastic differential
equations on nonholonomic manifolds when the fundamental Wiener processes
are associated to h-- and v--components and corresponding Euclidean carts.
The corresponding stochastic integrals are defined in the \ sense of
Stratonovich on any such open regions of \ $\mathbf{V}$ and $O(\mathbf{V})$
when the canonical realization of \ multidimensional Wiener processes is
used and $\delta \gamma ^{\alpha ^{\prime }}(\tau )\rightarrow \delta
\mathcal{W}^{\alpha ^{\prime }}.$  The stochastic differential
equation describing N--adapted diffusion on a nonholonomic manifold is%
\begin{eqnarray}
\delta u^{\alpha } &=&\mathbf{e}_{\ \alpha ^{\prime }}^{\alpha }(\tau )\circ
\delta \mathcal{W}^{\alpha ^{\prime }}+\mathbf{A}^{\alpha }(\tau )d\tau
\label{mathmodst} \\
\delta \mathbf{e}_{\ \alpha ^{\prime }}^{\alpha }(\tau ) &=&-\widehat{%
\mathbf{\Gamma }}_{\ \beta \nu }^{\alpha }(\tau )\mathbf{e}_{\ \alpha
^{\prime }}^{\nu }(\tau )\circ \delta u^{\beta },  \notag
\end{eqnarray}%
where the components of d--vector $\mathbf{A}^{\alpha }(\tau )=(A^{i}(\tau
),A^{a}(\tau ))$ are introduced additionally in order to model various \
types of stochastic processes and take into account possible external
forces. Here $\delta ^{\alpha ^{\prime }\beta ^{\prime }}\mathbf{e}_{\
\alpha ^{\prime }}^{\alpha }(\mathbf{u}(\tau ))\mathbf{e}_{\ \beta ^{\prime
}}^{\beta }(\mathbf{u}(\tau ))=\mathbf{g}^{\alpha \beta }$ with $\delta
^{\alpha ^{\prime }\beta ^{\prime }}$ taken as the flat Euclidean metric
splitting as $\delta ^{\alpha ^{\prime }\beta ^{\prime }}=\left( \delta
^{i^{\prime }j^{\prime }},\delta ^{a^{\prime }b^{\prime }}\right).$

Extending on $O(\mathbf{V})$ the definition of fundamental d--vector fields
from (\ref{astrat}), $\mathbf{L}_{\alpha ^{\prime }}\rightarrow \ ^{O}%
\mathbf{L}_{\alpha ^{\prime }}$ and $\mathbf{L}_{0}\rightarrow \ ^{O}\mathbf{%
L}_{0},$ where
$$\ ^{O}\mathbf{L}_{\alpha ^{\prime }} =\mathbf{e}_{\ \alpha ^{\prime
}}^{\alpha }\mathbf{e}_{\ \alpha }-\widehat{\mathbf{\Gamma }}_{\ \beta \nu
}^{\alpha }(u^{\mu })\mathbf{e}_{\ \alpha ^{\prime }}^{\beta }\mathbf{e}_{\
\beta ^{\prime }}^{\nu }\frac{\partial }{\mathbf{e}_{\ \beta ^{\prime
}}^{\beta }},  \ ^{O}\mathbf{L}_{0} =\mathbf{A}^{\alpha }(\tau ,\mathbf{U})\mathbf{e}_{\
\alpha }-\widehat{\mathbf{\Gamma }}_{\ \beta \nu }^{\alpha }\mathbf{A}%
^{\beta }\ \mathbf{e}_{\ \beta ^{\prime }}^{\nu }(\tau )\frac{\partial }{%
\mathbf{e}_{\ \beta ^{\prime }}^{\beta }},$$
we provide a nonholonomic generalization (horizontal lift of the diffusion
d--operator $\ ^{\mathbf{V}}\widetilde{\mathbf{A}}$) of the diffusion
generator $\widetilde{\mathbf{A}}$ on the bundle of orthonormalized
N--adapted frames,%
\begin{equation}
\ ^{O}\widetilde{\mathbf{A}}=\frac{\rho }{2}\sum\limits_{\alpha ^{\prime
}=1}^{n+m}\ ^{O}\mathbf{L}_{\alpha ^{\prime }}\ ^{O}\mathbf{L}_{\alpha
^{\prime }}+\ ^{O}\mathbf{L}_{0}.  \label{loper}
\end{equation}%
For any projection of a function $f$ in $O(\mathbf{V})$ to $\mathbf{V}$
(when, for instance, $\ f(\mathbf{r})=f(\mathbf{u},0),$ $\mathbf{r}%
=(u^{\alpha },\mathbf{e}_{\ \beta }^{\beta ^{\prime }})),$ we can write
\begin{equation*}
\ ^{O}\widetilde{\mathbf{A}}f(\mathbf{r})=\ ^{\mathbf{V}}\widetilde{\mathbf{A%
}}f(\mathbf{u}),
\end{equation*}%
\begin{eqnarray}
\mbox{ where \ } \ ^{\mathbf{V}}\widetilde{\mathbf{A}}&=&\frac{\rho }{2}\sum\limits_{\alpha
^{\prime }}\mathbf{e}_{\ \alpha ^{\prime }}^{\alpha }\mathbf{e}_{\ \alpha }(%
\mathbf{e}_{\ \alpha ^{\prime }}^{\beta }\mathbf{e}_{\ \beta })+\mathbf{A}%
^{\beta }\ \mathbf{e}_{\ \beta }=\frac{\rho }{2}\widehat{\bigtriangleup }+%
\mathbf{A}^{\beta }\ \mathbf{e}_{\ \beta }  \label{diffusd} \\
\mbox{ and\ }
\widehat{\bigtriangleup }&=&\frac{1}{2}\mathbf{g}^{\alpha \beta }\left[
\mathbf{e}_{\ \alpha }\mathbf{e}_{\ \beta }+\mathbf{e}_{\ \beta }\mathbf{e}%
_{\ \alpha }+\left( \widehat{\mathbf{\Gamma }}_{\ \alpha \beta }^{\nu }+%
\widehat{\mathbf{\Gamma }}_{\ \beta \alpha }^{\nu }\right) \mathbf{e}_{\ \nu
}\right]  \label{dlb}
\end{eqnarray}%
is the canonical Laplace--Beltrami d--operator defined by the inverse
coefficients $\mathbf{g}^{\alpha \beta }$ of d--metric (\ref{dm}), \ the
canonical d--connection $\widehat{\mathbf{\Gamma }}_{\ \alpha \beta }^{\nu }$
(\ref{candcon}) and N--connection $\mathbf{N}=\{N_{k}^{a}\}$ (\ref{whitney}%
).

The operators $\ ^{\mathbf{V}}\widetilde{\mathbf{A}}$ (\ref{diffusd}) and $%
\widehat{\bigtriangleup }$ (\ref{dlb}) allows us to formulate a generalized
Kolmogorov backward equation on a nonholonomic manifold $\mathbf{V,}$%
\begin{eqnarray}
\partial _{\tau }\underline{f}(\tau ,\mathbf{u}) &=&\ ^{\mathbf{V}}%
\widetilde{\mathbf{A}}\underline{f}(\tau ,\mathbf{u}),  \label{kfpnh} \\
\underline{f}(0,\mathbf{u}) &=&f(\mathbf{u}).  \notag
\end{eqnarray}%
Since the canonical Laplace--Beltrami d--operator is self--adjoint, $%
\widehat{\bigtriangleup }=\ ^{\ast }\widehat{\bigtriangleup },$ we can
construct a self--adjoint $\ ^{\ast }\left( ^{\mathbf{V}}\widetilde{\mathbf{A%
}}\right) $ to $\ ^{\mathbf{V}}\widetilde{\mathbf{A}}$ as we \ explained
above in footnote \ref{fn1}. As a result, the corresponding generalized
Fokker--Planck equation on \ $\mathbf{V}$ is
\begin{equation}
\partial _{\tau }\digamma =-\frac{1}{\sqrt{|\mathbf{g}_{\alpha \beta }|}}%
\mathbf{e}_{\ \nu }(\sqrt{|\mathbf{g}_{\alpha \beta }|}\mathbf{A}^{\nu
}\digamma )+\frac{\rho }{2}\widehat{\bigtriangleup }\digamma ,  \label{fpnh}
\end{equation}%
where $\digamma =\digamma (\tau ,\ ^{1}\mathbf{u;}0\mathbf{,\ }^{2}\mathbf{u}%
)$ is the transition probability with the initial condition $\digamma (0,\
^{1}\mathbf{u;}0\mathbf{,\ }^{2}\mathbf{u})=\delta (\ ^{1}\mathbf{u-\ }^{2}%
\mathbf{u})$ for any two points $\ ^{1}\mathbf{u,\ }^{2}\mathbf{u\in }$\ $%
\mathbf{V}$ and adequate \ boundary conditions at infinity.\footnote{%
The same equations can be considered for the probability density $\varphi
(\tau ,\mathbf{u})$ but for the initial condition $\varphi (\tau =0,\mathbf{u%
})=\ ^{0}\varphi (\mathbf{u}).$} Finally we note that we get usual
evolution/diffusion equations in nonholonomic curved spaces, \ $\partial
_{\tau }\digamma =\frac{\rho }{2}\widehat{\bigtriangleup }\digamma ,$ if we
impose the condition that the divergence in (\ref{fpnh}) (i.e. the first
term in the right part of equation) is taken zero.

\section{Nonholonomic Diffusion in General Relativity}

\label{snrdife}

In the derivation of relativistic diffusion equations and constructing
gravity theories, we have to take into account in an appropriate way the
fact that the speed of light has a constant maximal value. Geometrically,
such a fundamental experimental fact is encoded into the special theory of
relativity as the condition that the velocity space is a hyperboloid (i.e. a
special type three dimensional, 3--d, Riemannian manifold) embedded into the
4--d velocity Minkowski space.  A formal
geometric analogy between Euclidian/Riemannian/Finsler etc geometries and
respective ones with ''pseudo'' signatures can be preserved by introducing a
formal ''imaginary'' time like in the ''early'' works on general relativity %
\cite{landau,moll}.

In this paper, we elaborate a geometric and stochastic formalism in order to
include in the scheme the relativistic diffusion processes with exact
solutions for gravitational field equations. Such solutions can be
constructed in general form only by imposing corresponding nonholonomic
constraints on the systems of partial differential equations for dynamical
and/or stochastic systems. The goal of this section is to consider an
extension of the theory of relativistic diffusion on (flat) Minkowski and
(curved) Einstein spaces when such spacetimes are enabled with conventional
h-- and v--splitting (respectively, with trivial and/or nontrivial
N--connection structure). The values of coefficients metrics and connections
are considered to be given from certain (necessary smooth class) solutions
of classical gravitational and matter fields equations. In the next section %
\ref{seses}, the scheme will be completed by elaborating a method of
generating both classical and stochastic solutions of \ Einstein equations.

\subsection{The special relativistic nonholonomic diffusion}

\label{ssnrd}The velocity space in special relativity is characterized by a
noncompact hyperbolic structure which for the 4--d Minkowski spacetime $\
_{1}^{3}M$ is parametr\-iz\-ed by a corresponding relation
\begin{equation}
-(v^{1})^{2}+(v^{2})^{2}+(v^{3})^{2}+(v^{4})^{2}=-1,  \label{hypercond}
\end{equation}%
where $v^{\alpha }$ are normalized velocity variables defined for the
typical fiber of tangent bundle $T(\ _{1}^{3}M).$ We can introduce the
laboratory time $t=\tau u^{1}/c,$ where $c$ is the light velocity and $\tau $
is the proper time, and express $v^{1}=\left[
1+(v^{2})^{2}+(v^{3})^{2}+(v^{4})^{2}\right] ^{1/2}.$ The local coordinates
on $T(\ _{1}^{3}M)$ can be para\-metrized in the form $u^{\alpha
}=(x^{i},y^{a}=v^{a}),$ where $i,j,...=1,2,3,4$ and $a,b,c...=5,6,7,8.$ It
is considered a conventional 1+3 splitting for $\ _{1}^{3}M.$ We shall write
$\ ^{h}T(\ _{1}^{3}M)$ if the v--coordinates are subjected to constraints of
type (\ref{hypercond}) and say that its typical fiber space is a hyperbolic
velocity space (when the hyperboloid is embedded into the 4--d Minkovski
spacetime). Such a space is naturally enabled with a corresponding metric, $%
h_{\widehat{a}\widehat{b}}(v^{c}),$ and linear connection (Christoffel
connection coefficients on the hyperboloid), $\gamma _{\widehat{b}\widehat{c}%
}^{\widehat{a}}(v^{e}),$ when
\begin{equation}
h_{\widehat{a}\widehat{b}}(v^{c})=\delta _{\widehat{a}\widehat{b}}-v^{%
\widehat{a}}v^{\widehat{b}}/(v^{1})^{2},\ \gamma _{\widehat{b}\widehat{c}}^{%
\widehat{a}}(v^{e})=v^{\widehat{a}}h_{\widehat{b}\widehat{c}},  \label{aux2}
\end{equation}%
where (in this subsection) $\widehat{a},\widehat{b},...=2,3,4.$ We can
prescribe additionally any nonholonomic $2+2$ and/or $4+4$ splitting on $\
_{1}^{3}M$ and/or $T(\ _{1}^{3}M)$ by prescribing on such spaces
corresponding nonholonomic distributions with associated N--connection
structures of type (\ref{whitney}). A nonholonomic distribution in $T(\
_{1}^{3}M)$ is given by $\mathbf{N}=\{N_{k}^{\widehat{a}}(u^{\widehat{\alpha
}})\},$ local coordinates $\widehat{\mathbf{u}}=\{u^{\widehat{\alpha }%
}=(x^{i},v^{\widehat{a}})\}.$ The d--metric structure $\ \widehat{\mathbf{g}}
$ on $T(\ _{1}^{3}M)$) can be written (in a form similar to (\ref{dm}))
\begin{equation}
\ \widehat{\mathbf{g}}\mathbf{=}\eta _{ij}{dx^{i}\otimes dx^{j}}+h_{\widehat{%
a}\widehat{b}}(dv^{\widehat{a}}+N_{k}^{\widehat{a}}dx^{k}){\otimes }(dv^{%
\widehat{b}}+N_{k}^{\widehat{b}}dx^{k}),  \label{dmms}
\end{equation}%
where $\eta _{ij}=(-,+,+,+),$ for which, using formulas (\ref{candcon}), we
can compute the corresponding coefficients of the canonical d--connection $%
\widehat{\mathbf{\Gamma }}_{\ \widehat{\alpha }\widehat{\beta }}^{\widehat{%
\gamma }}=(\widehat{L}_{jk}^{i},\widehat{L}_{\widehat{b}k}^{\widehat{a}},$ $%
\widehat{C}_{j\widehat{c}}^{i},\widehat{C}_{\widehat{b}\widehat{c}}^{%
\widehat{a}}).$

For nonholonomic distributions with $N_{k}^{\widehat{a}}(u^{\widehat{\alpha }%
})$ when $\Omega _{\ ji}^{\widehat{a}}=0,$ see formulas (\ref{dtors}), we
get $\widehat{C}_{\widehat{b}\widehat{c}}^{\widehat{a}}=\gamma _{\widehat{b}%
\widehat{c}}^{\widehat{a}}(v^{e}).$ But, in general, we work with arbitrary
frame and coordinate transforms and the fundamental geometric objects on $\
^{h}T(\ _{1}^{3}M)$ with nontrivial h- and v--splitting induced by $\mathbf{%
N,}$ when $\ \widehat{\mathbf{g}}=\{\ \widehat{\mathbf{g}}_{\widehat{\alpha }%
\widehat{\beta }}=[\widehat{g}_{ij}(u^{\widehat{\alpha }}),h_{\widehat{a}%
\widehat{b}}(u^{\widehat{\alpha }})]\}.$ It is possible to introduce
N--adapted orthonormalized frame bases $\mathbf{E}_{\widehat{\alpha }%
^{\prime }}=\mathbf{E}_{\ \widehat{\alpha }^{\prime }}^{\underline{\widehat{%
\alpha }}}(\widehat{\mathbf{u}})\partial _{\underline{\widehat{\alpha }}}=%
\mathbf{E}_{\ \widehat{\alpha }^{\prime }}^{\widehat{\alpha }}(\widehat{%
\mathbf{u}})\mathbf{E}_{\widehat{\alpha }}$ for which
\begin{eqnarray}
\ \widehat{\mathbf{g}}_{\widehat{\alpha }^{\prime }\widehat{\beta }^{\prime
}} &=&[\widehat{g}_{i^{\prime }j^{\prime }}=\eta _{i^{\prime }j^{\prime
}},h_{\widehat{a}^{\prime }\widehat{b}^{\prime }}=\delta _{\widehat{a}%
^{\prime }\widehat{b}^{\prime }}]=\mathbf{E}_{\ \widehat{\alpha }^{\prime
}}^{\widehat{\alpha }}\mathbf{E}_{\ \widehat{\beta }^{\prime }}^{\widehat{%
\beta }}\ \widehat{\mathbf{g}}_{\widehat{\alpha }\widehat{\beta }},
\label{frtr2} \\
\widehat{g}_{i^{\prime }j^{\prime }} &=&E_{\ i^{\prime }}^{i}E_{\ j^{\prime
}}^{j}\ \widehat{g}_{ij},\ h_{\widehat{a}^{\prime }\widehat{b}^{\prime
}}=E_{\ \widehat{a}^{\prime }}^{\widehat{a}}E_{\ \widehat{b}^{\prime }}^{%
\widehat{b}}\ h_{\widehat{a}\widehat{b}},  \notag
\end{eqnarray}%
when the values with ''un--primed'' indices are given by the coefficients of
d--metric (\ref{dmms}) and (\ref{aux2}). In our approach, the \ frame
transform are written with capital letters, $\mathbf{E}_{\ \widehat{\alpha }%
^{\prime }}^{\underline{\widehat{\alpha }}}=[E_{\ i^{\prime }}^{i},E_{\
\widehat{a}^{\prime }}^{\widehat{a}}],$ for spaces with pseudo--Euclidian
and/or hyperbolic fiber/velocity space. The values $E_{\ i^{\prime }}^{i}$
parametrize moving frames adapted to the position spacetime $\ _{1}^{3}M$
and the values $E_{\ \widehat{a}^{\prime }}^{\widehat{a}}(\tau )$ state
moving velocity frames adapted to the typical \ fiber of \ $\ ^{h}T(\
_{1}^{3}M).$

Working with ''hyperbolic geometric data'' $\ \widehat{\mathbf{g}}$ (\ref%
{dmms}), $\widehat{\mathbf{\Gamma }}_{\ \widehat{\alpha }\widehat{\beta }}^{%
\widehat{\gamma }}$ and $\mathbf{E}_{\ \widehat{\alpha }^{\prime }}^{%
\underline{\widehat{\alpha }}}$ (\ref{frtr2}) for a prescribed N--connection
structure $\mathbf{N:\ }\ TT(\ _{1}^{3}M)=hT(\ _{1}^{3}M)\oplus vT(\
_{1}^{3}M),$ we can define a model of nonholonomic relativistic diffusion in
special relativity (i.e. on total space $T(\ _{1}^{3}M)$) similarly to the
constructions provided in section \ref{ssdnhm}. Let us consider the frame
bundle space $\mathbf{F}(\ ^{h}T(\ _{1}^{3}M))$ with local coordinates $%
\widehat{\mathbf{r}}=\{u^{\widehat{\alpha }},\mathbf{E}_{\ \widehat{\alpha }%
^{\prime }}^{\underline{\widehat{\alpha }}}\}$ and identify $dx^{i}(\tau
)=v^{i}(\tau )d\tau ,$ where $\tau $ can be interpreted as an evolution
parameter along the world lines of the particles which can be chosen as the
proper time (we can take $\tau $ as the temperature in gravitational
thermo--field models etc). The N--adapted relativistic stochastic equations
on $\ ^{h}T(\ _{1}^{3}M)$ are similar to (\ref{mathmodst}),

\begin{eqnarray}
\delta u^{\widehat{\alpha }} &=&\mathbf{E}_{\ \widehat{\alpha }^{\prime }}^{%
\widehat{\alpha }}(\tau )\circ \delta \mathcal{W}^{\widehat{\alpha }^{\prime
}}+\mathbf{A}^{\widehat{\alpha }}(\tau )d\tau  \label{relstde} \\
\delta \mathbf{E}_{\ \widehat{\alpha }^{\prime }}^{\widehat{\alpha }}(\tau )
&=&-\widehat{\mathbf{\Gamma }}_{\ \widehat{\beta }\widehat{\nu }}^{\widehat{%
\alpha }}(\tau )\mathbf{E}_{\ \widehat{\alpha }^{\prime }}^{\widehat{\nu }%
}(\tau )\circ \delta u^{\widehat{\beta }},  \notag
\end{eqnarray}%
when the velocity coordinates are subjected to the hyperbolicity conditions (%
\ref{hypercond}) and the components of drift d--vector are parametrized $%
\mathbf{A}^{\widehat{\alpha }}(\tau )=(A^{i}(\tau )=b^{i}(\tau ),A^{\widehat{%
a}}(\tau )=B^{\widehat{a}}(\tau )).$

The stochastic differential equations (\ref{relstde}) adapted to the
nonholonomic hyperbolic velocity structure split into two families
(respectively for the position coordinates and for the velocity type
coordinates),%
\begin{eqnarray}
\delta x^{i}(\tau ) &=&E_{\ i^{\prime }}^{i}(\tau )\circ \delta \mathcal{W}%
^{i^{\prime }}+b^{i}(\tau )d\tau  \label{pdiff} \\
dE_{\ i^{\prime }}^{i}(\tau ) &=&-\widehat{L}_{jk}^{i}(\tau )E_{\ i^{\prime
}}^{k}(\tau )\circ dx^{k},  \notag
\end{eqnarray}%
and
\begin{eqnarray}
\delta v^{\widehat{a}} &=&E_{\ \widehat{a}^{\prime }}^{\widehat{a}}(\tau
)\circ \delta \mathcal{W}^{\widehat{a}^{\prime }}+B^{\widehat{a}}(\tau )d\tau
\label{fpdiff} \\
\delta E_{\ \widehat{a}^{\prime }}^{\widehat{a}}(\tau ) &=&-\widehat{C}_{%
\widehat{b}\widehat{c}}^{\widehat{a}}(\tau )E_{\ \widehat{a}^{\prime }}^{%
\widehat{c}}(\tau )\circ \delta v^{\widehat{b}}.  \notag
\end{eqnarray}%
where a nontrivial $B^{\widehat{a}}(\tau )=F^{\widehat{a}}/m_{0}$ is defined
by spacial components of a 4--force $F^{a}$ acting on particles of rest mass
$m_{0}.$ For trivial N--connection structure, the equations (\ref{pdiff})
and (\ref{fpdiff}) transform respectively into relativistic stochastic
equations (2) and (3) proposed in Ref. \cite{herrmann1}. In our approach, we
can introduce a nonholonomic dynamics in the velocity space of special
relativity, via corresponding N--connection structure and generalized
d--connections. This way we can model various theories with
restricted/brocken Lorentz symmetry etc.

For simplicity, we omit here the considerations when from (\ref{relstde}) a
relativistic theory of $(\mathbf{A,L})$--diffusion is derived. \footnote{%
Corresponding formulas are similar to (\ref{diffusd})--(\ref{fpnh}) with
that difference that the ''hyperbolic'' small Greek and velocity indices,
for \ this section, are with ''hats'' emphasizing the fact that the
nonholonomic diffusion evolution is adapted to the condition of constant
speed of light.}

\subsection{Nonholonomic diffusion and gravitational interactions}

There are two classes of relativistic diffusion theories for gravity:

\begin{enumerate}
\item The first one is for modelling relativistic stochastic processes on
fixed classical curved spacetimes. We have to consider, instead of the flat
Minkovski spacetime $\ _{1}^{3}M$ and metric $\ \widehat{\mathbf{g}}$ (\ref%
{dmms}), a (pseudo) Riemannian spacetime $\mathbf{V}$ and a solution $\
\mathbf{g}_{\mu \nu }$ in general relativity (GR). The four--velocity $%
v^{\mu }$ of massive particles in GR satisfies the condition
\begin{equation}
\mathbf{g}_{\mu \nu }(u^{\alpha })v^{\mu }v^{\nu }=-1,  \label{hip3}
\end{equation}%
(here $\mu ,\nu =1,2,3,4$), which is \ a generalization of (\ref{hypercond})
(we suppose that in any point $\mathbf{u}\in \mathbf{V}$ the hyperbolicity
condition holds in the typical fiber of $T\mathbf{V)}.$ By using the
orthonormalized moving frames of the pseudo--Riemannian manifold we get the
same formulas as for relativistic diffusion in special relativity, see
details in Section 3 of \cite{herrmann1}. If the Levi--Civita connection $%
\nabla $ is changed into the canonical d--connection $\widehat{\mathbf{D}}$
(or, for other models, $\mathbf{D}=\ ^{c}\mathbf{D,}$ for instance, the
Cartan d--connection), we get relativistic models on nonholonomic (pseudo)
Riemann and/or Lagrange--Finsler spaces  \cite%
{vstoch5,vstoch6}.

\item The second class of theories is that when d--metrics $\mathbf{g}=\{%
\mathbf{g}_{\mu \nu }\}$ (\ref{dm}) are exact solutions of Einstein
equations in a gravity theory for $\widehat{\mathbf{D}},$ or its restriction
to $\nabla $ (with $Z_{\ \alpha \beta }^{\gamma }=0$ in formulas (\ref{deflc}%
)), when some coefficients $\mathbf{g}_{\mu \nu }$ take stochastic values,
which results in a relativistic gravitational diffusion of metrics in
general relativity (for various generalizations). We work with nonholonomic
gravitational configurations with associated N--connection splitting because
in such cases we are able to solve the Einstein equations in general forms
and analyze mutual diffusions of metrics (the splitting of equations is not
possible for not N--adapted constructions with the Levi--Civita connection,
see   section \ref{seses}).
\end{enumerate}

In this section we shall elaborate a model of relativistic diffusion for a
nonholonomic (pseudo) Riemannian spaces $\mathbf{V}$ with prescribed
N--connection structure $N:\ $ $T\mathbf{V=}h\mathbf{V\oplus }v\mathbf{V,}$
when $\dim \mathbf{V}=4.$ The constructions are performed \ for the
canonical d--connection $\widehat{\mathbf{D}}$ and d--metrics being solution
of the nonholonomic Einstein equations with limits of type $\widehat{\mathbf{%
D}}_{\mid Z\rightarrow 0}\rightarrow \nabla $.

\subsubsection{N--adapted frames and stochastic equations in GR}

Orthonormalized N--adapted frames on $\mathbf{V,}$ $\mathbf{e}_{\alpha
^{\prime }}=\mathbf{e}_{\ \alpha ^{\prime }}^{\underline{\alpha }}(\mathbf{u}%
)\partial _{\underline{\alpha }}=\mathbf{e}_{\ \alpha ^{\prime }}^{\alpha }(%
\mathbf{u})\mathbf{e}_{\alpha },$ with $\mathbf{e}_{\alpha }$ being of type (%
\ref{ddr}), can be defined by any nondegernerated matrix fields $\mathbf{e}%
_{\ \alpha ^{\prime }}^{\underline{\alpha }}(\mathbf{u})$ and $\mathbf{e}_{\
\alpha ^{\prime }}^{\alpha }(\mathbf{u})$ $\ ($and/or, respectively, their
inverses, $\mathbf{e}_{\underline{\alpha }\ }^{\ \alpha ^{\prime }}(\mathbf{u%
})$ and $\mathbf{e}_{\alpha \ }^{\ \alpha ^{\prime }}(\mathbf{u}))$
subjected to the conditions%
\begin{equation}
\eta _{\alpha ^{\prime }\beta ^{\prime }}=\mathbf{g}_{\alpha \beta }\mathbf{e%
}_{\ \alpha ^{\prime }}^{\alpha }(\mathbf{u})\ \mathbf{e}_{\ \beta ^{\prime
}}^{\beta }(\mathbf{u}),  \label{hip4}
\end{equation}%
where $\eta _{\alpha ^{\prime }\beta ^{\prime }}=(1,1,-1,1),$ and the
space--time orientation of coordinates is chosen in a form which will be
convenient for constructing exact generic off--diagonal solutions of
Einstein equations (see next section).

We can compute the orthonormalized components $v^{\alpha ^{\prime }}$ of a
4--vector $v^{\alpha },$ $v^{\alpha ^{\prime }}=\mathbf{e}_{\alpha \ }^{\
\alpha ^{\prime }}(\mathbf{u})v^{\alpha },$ when $\eta _{\alpha ^{\prime
}\beta ^{\prime }}v^{\alpha ^{\prime }}v^{\beta ^{\prime }}=-1$ which can be
proven using algebraic relations (\ref{hip3}) and (\ref{hip4}). By
definition, $\widehat{\mathbf{D}}$ is metric compatible and we can impose
the condition $\widehat{\mathbf{D}}_{\beta }\mathbf{e}_{\ \alpha ^{\prime
}}^{\alpha }=0$ and $\widehat{\mathbf{D}}_{\beta }\mathbf{e}_{\ \alpha
^{\prime }}^{\alpha }=0.$ We get such a transformation law for the canonical
d--connection coefficients,%
\begin{equation*}
\widehat{\mathbf{\Gamma }}_{\ \beta \nu ^{\prime }}^{\alpha ^{\prime }}(%
\mathbf{u})=\mathbf{e}_{\alpha \ }^{\ \alpha ^{\prime }}(\mathbf{u})\left(
\widehat{\mathbf{\Gamma }}_{\ \beta \nu }^{\alpha }\mathbf{e}_{\ \nu
^{\prime }}^{\nu }(\mathbf{u})+\mathbf{e}_{\beta }\mathbf{e}_{\ \nu ^{\prime
}}^{\alpha }(\mathbf{u})\right) ,
\end{equation*}%
which can be decomposed into h-- and v--components using spliting of type $%
\mathbf{e}_{\alpha \ }^{\ \alpha ^{\prime }}=\left( e_{i\ }^{\ i^{\prime
}},e_{a\ }^{\ a^{\prime }}\right) ,\mathbf{e}_{\beta }=\left( \mathbf{e}%
_{j},e_{b}\right) $ and $\widehat{\mathbf{\Gamma }}_{\ \alpha \beta
}^{\gamma }=(\widehat{L}_{jk}^{i},\widehat{L}_{bk}^{a},\widehat{C}_{jc}^{i},%
\widehat{C}_{bc}^{a}).$ The values $\widehat{\mathbf{\Gamma }}_{\ \beta \nu
^{\prime }}^{\alpha ^{\prime }}$ are analogs of spin connection coefficients
$\Gamma _{\ \beta \nu ^{\prime }}^{\alpha ^{\prime }}$ in general
relativity; $\widehat{\mathbf{\Gamma }}_{\ \beta \nu ^{\prime }\mid
Z\rightarrow 0}^{\alpha ^{\prime }}\rightarrow \Gamma _{\ \beta \nu ^{\prime
}}^{\alpha ^{\prime }}$.

The N--adapted parallel transport of a d--vector $v^{\alpha ^{\prime }}$
with respect to an orthonormalized N--frame $\mathbf{e}_{\beta ^{\prime }}$
is defined by line elements
\begin{equation*}
\delta v^{\alpha ^{\prime }}=-\widehat{\mathbf{\Gamma }}_{\ \beta \gamma
^{\prime }}^{\alpha ^{\prime }}(u^{\beta })v^{\gamma ^{\prime }}\delta
u^{\beta },\mbox{ \ where \ }\delta u^{\beta }=\mathbf{e}_{\ \beta ^{\prime
}}^{\beta }(u^{\alpha })v^{\beta ^{\prime }}\delta \tau .
\end{equation*}%
We can say that the change of spacial components (labeled by small Greeck
indices with hats, for a 3--d velocity space which in the relativistic case
is subjected to the condition of hyperbolicity (\ref{hypercond})) of the
velocity vector d--field $v^{\widehat{\alpha }^{\prime }}$ is driven by a
formal (gravitational) force $\widehat{\mathbf{\Gamma }}_{\ \beta \widehat{%
\gamma }^{\prime }}^{\widehat{\alpha }^{\prime }}(u^{\beta }).$ One could be
additional contributions from a ''stochastic force'' (such a noise can from
any classical or quantum gravitational, or matter fields, fluctuations)
associated to a Wiener process $\delta \mathcal{W}^{\widehat{\alpha }%
^{\prime }}.$ For relativistic constructions, we have to consider stochastic
processes along the orthonormal frames $E_{\widehat{\alpha }}^{\widehat{%
\alpha }^{\prime }}(v^{\beta })$ as we considered in (\ref{frtr2}). In this
subsection, the position space is not the Minkovski spacetime $\ _{1}^{3}M$
but a nonholonomic (pseudo) Riemannian one $\mathbf{V;}$ we can generalize
the constructions considering that in any point $\mathbf{u}\in \mathbf{V}$
it is defined a hyperbolic velocity space with the ''fiber'' metric and
connections determined by
\begin{equation*}
h_{\widehat{\alpha }\widehat{\beta }}(v^{\gamma })=\delta _{\widehat{\alpha }%
\widehat{\beta }}-v^{\widehat{\alpha }}v^{\widehat{\beta }}/(v^{3})^{2},\
\gamma _{\widehat{\beta }\widehat{\gamma }}^{\widehat{\alpha }}(v^{\varphi
})=v^{\widehat{\alpha }}h_{\widehat{\alpha }\widehat{\beta }}
\end{equation*}
as in (\ref{aux2}) but with that difference that in this subsection indices
run different values (for instance, $\widehat{\alpha }=1,2,4)$ than those in
used in subsection \ref{ssnrd}.

Considering that on a nonholonomic (pseudo) Riemannian manifold the Wiener
process is moved along the orthonormalized frames $E_{\widehat{\alpha }}^{%
\widehat{\alpha }^{\prime }}(v^{\beta })$ in the 3--d hyperbolic velocity
space on $u^{\beta }\in \mathbf{V,}$ when (with summation on repeating
indices)
\begin{equation}
E_{\widehat{\alpha }^{\prime }}^{\widehat{\alpha }}E_{\widehat{\alpha }%
^{\prime }}^{\widehat{\beta }}=h^{\widehat{\alpha }\widehat{\beta }},%
\mbox{
\ equivalently \ }h_{\widehat{\alpha }\widehat{\beta }}E_{\widehat{\alpha }%
^{\prime }}^{\widehat{\alpha }}E_{\widehat{\beta }^{\prime }}^{\widehat{%
\beta }}=\delta _{\widehat{\alpha }^{\prime }\widehat{\beta }^{\prime }},
\label{algrel}
\end{equation}%
where $h^{\widehat{\alpha }\widehat{\beta }}$ is inverse to the hyperbolic
metric $h_{\widehat{\alpha }\widehat{\beta }},$ the infinitesimal motion of
the velocity $v^{\widehat{\alpha }}$ is determined by equations%
\begin{equation}
\delta v^{\widehat{\alpha }}=E_{\widehat{\alpha }^{\prime }}^{\widehat{%
\alpha }}\delta v^{\widehat{\alpha }^{\prime }}\mbox{ \ and \ }\delta E_{%
\widehat{\alpha }^{\prime }}^{\widehat{\alpha }}(\tau )=-\gamma _{\widehat{%
\beta }\widehat{\gamma }}^{\widehat{\alpha }}(v^{\widehat{\varphi }})E_{%
\widehat{\alpha }^{\prime }}^{\widehat{\gamma }}\delta v^{\widehat{\beta }}.
\label{pdiffg}
\end{equation}

A random N--adapted curve on $\mathbf{V,}$ parametrized in the phase--space
can be introduced as a Wiener process: $dv^{\widehat{\alpha }^{\prime
}}\rightarrow \delta \mathcal{W}^{\widehat{\alpha }^{\prime }}.$ We can
consider a formal noise force $\ ^{noise}B^{\widehat{\alpha }}=E_{\widehat{%
\alpha }^{\prime }}^{\widehat{\alpha }}(\tau )\circ \delta \mathcal{W}^{%
\widehat{\alpha }^{\prime }}.$ Following constraints (\ref{hip4}) and
algebraic relations (\ref{algrel}), we conclude that the relativistic
diffusion on spacetime $\mathbf{V}$ is defined by a more restricted system
of coefficients because there are admitted only such $E_{\widehat{\alpha }%
^{\prime }}^{\widehat{\alpha }}(\tau )$ when a direct relation to the
hyperbolic geometry is established. We can say that we model via $\widehat{%
\mathbf{\Gamma }}_{\ \beta \nu ^{\prime }}^{\alpha ^{\prime }}$ a consistent
description of Markovian diffusion in general relativity but keeping certain
''geometric memory'' on nonholonomic h-v--splitting and spacetime (pseudo)
Riemannian geometry, all encoded into a corresponding Stratonovich
relativistic calculus.

The stochastic differential equations describing the N--adapted relativistic
diffusion of gravitational and external force fields\footnote{%
i.e. the nonholonomic Langevin equations in general relativity} are written
\begin{eqnarray}
\delta u^{\alpha } &=&e_{\ \alpha ^{\prime }}^{\alpha }(u^{\beta })v^{\alpha
^{\prime }}\delta \tau ,  \label{fpdiffg} \\
\delta v^{\widehat{\alpha }} &=&E_{\widehat{\alpha }^{\prime }}^{\widehat{%
\alpha }}(\tau )\circ \delta \mathcal{W}^{\widehat{\alpha }^{\prime }}-%
\widehat{\mathbf{\Gamma }}_{\ \beta \gamma ^{\prime }}^{\widehat{\alpha }%
}(u^{\gamma })e_{\ \alpha ^{\prime }}^{\beta }(u^{\gamma })v^{\gamma
^{\prime }}v^{\alpha ^{\prime }}\delta \tau +\ ^{ex}B^{\widehat{\alpha }%
}\delta \tau  \notag \\
\delta E_{\widehat{\alpha }^{\prime }}^{\widehat{\alpha }}(\tau ) &=&-\gamma
_{\widehat{\beta }\widehat{\gamma }}^{\widehat{\alpha }}(v^{\widehat{\varphi
}})E_{\widehat{\alpha }^{\prime }}^{\widehat{\gamma }}\circ \delta v^{%
\widehat{\beta }},  \notag
\end{eqnarray}%
where a possible additional external force $\ ^{ex}B^{\widehat{\alpha }}=F^{%
\widehat{a}}/m_{0}$ is defined by spacial components of a 4--force $F^{a}$
acting on particles of rest mass $m_{0}$ and $\tau $ is a parameter defined
along of world line of particles (in our case, moving with nonholonomic
constraints on $\mathbf{V}$); we can consider $\tau $ as the phase--space
proper time.

The system of stochastic equations (\ref{pdiffg}) and (\ref{fpdiffg}) is a
respective analogous of (\ref{pdiff}) and (\ref{fpdiff}). There are two
substantial difference between such systems of equations: the first one is
for general relativity when the N--connection structure is defined for the
base spacetime manifold but the second one is for special relativity
nonholonomically extended as a stochastic geometric model for $\mathbf{N:\ }%
\ TT(\ _{1}^{3}M)=hT(\ _{1}^{3}M)\oplus vT(\ _{1}^{3}M).$ Finally, we note
that, for instance, there are satisfied sufficient and necessary conditions
for the existence and uniqueness of N--adapted relativistic stochastic
differential equations (\ref{fpdiffg}) \ if the drift and diffusion
coefficients are subjected to uniform Lipschiz conditions (see details in %
\cite{elw,ikeda,emery}) and the stochastic process $\mathbf{X}(\tau )=\{%
\mathbf{u}(\tau ),\mathbf{v}(\tau )\}$ is N--adapted to the Wiener process $%
\mathcal{W}^{\widehat{\alpha }^{\prime }}(\tau ),$ when the output $\mathbf{X%
}(\ ^{2}\tau )$ is a function of $\mathcal{W}^{\widehat{\alpha }^{\prime
}}(\ ^{1}\tau )$ up to that time, for $\ ^{1}\tau \leq \ ^{2}\tau .$

\subsubsection{Diffusion and N--adapted stochastic GR processes}

A model of general relativistic of $(\mathbf{A,L})$--diffusion on the fiber
bundle $\mathbf{F}(\mathbf{V})$ \ with local coordinates $\ ^{E}\mathbf{r}%
=\{u^{\alpha }=(x^{i},y^{a}),v^{\widehat{\beta }},E_{\widehat{\alpha }%
^{\prime }}^{\widehat{\alpha }}\}$ can be derived from (\ref{pdiffg}) and (%
\ref{fpdiffg}). The N--adapted diffusion operator $\ ^{\mathbf{F(V)}}\mathbf{%
A}$ can be construction similarly to (\ref{loper}) by using operators
\begin{eqnarray*}
\ \ ^{F}L_{\widehat{a}^{\prime }} &=&E_{\widehat{\alpha }^{\prime }}^{%
\widehat{\alpha }}\frac{\partial }{\partial v^{\widehat{\alpha }}}-\gamma _{%
\widehat{\varepsilon }\widehat{\gamma }}^{\widehat{\alpha }}(v^{\widehat{%
\varphi }})E_{\widehat{\alpha }^{\prime }}^{\widehat{\gamma }}E_{\widehat{%
\beta }^{\prime }}^{\widehat{\varepsilon }}\frac{\partial }{\partial E_{%
\widehat{\beta }^{\prime }}^{\widehat{\alpha }}}, \\
\ \ ^{F}\mathbf{L}_{0} &=&\mathbf{e}_{\ \alpha ^{\prime }}^{\alpha
}(u^{\beta })v^{\alpha ^{\prime }}\mathbf{e}_{\alpha }-\widehat{\mathbf{%
\Gamma }}_{\ \beta \gamma ^{\prime }}^{\widehat{\alpha }}(u^{\gamma })%
\mathbf{e}_{\ \alpha ^{\prime }}^{\beta }(u^{\gamma })v^{\gamma ^{\prime
}}v^{\alpha ^{\prime }}\frac{\partial }{\partial v^{\widehat{\alpha }}} +\ ^{ex}B^{\widehat{\alpha }}\frac{\partial }{\partial v^{\widehat{\alpha }%
}}-\gamma _{\widehat{\varepsilon }\widehat{\gamma }}^{\widehat{\alpha }}(v^{%
\widehat{\varphi }})E_{\widehat{\beta }^{\prime }}^{\widehat{\gamma }}B^{%
\widehat{\varepsilon }}\frac{\partial }{\partial E_{\widehat{\beta }^{\prime
}}^{\widehat{\alpha }}},
\end{eqnarray*}%
where the force $B^{\widehat{\varepsilon }}$ consists respectively from the
gravitational and external components, $$B^{\widehat{\varepsilon }}=-\widehat{%
\mathbf{\Gamma }}_{\ \beta \gamma ^{\prime }}^{\widehat{\varepsilon }%
}(u^{\gamma })e_{\ \alpha ^{\prime }}^{\beta }(u^{\gamma })v^{\gamma
^{\prime }}v^{\alpha ^{\prime }}+\ ^{ex}B^{\widehat{\varepsilon }}.$$ For
real applications, it is convenient to use the operator $\ ^{\mathbf{P}}%
\mathbf{A}$ as the projection of $\ ^{\mathbf{F(V)}}\mathbf{A}$ on the phase
space with coordinates $\mathbf{r}=\{u^{\alpha }=(x^{i},y^{a}),v^{\widehat{%
\beta }}\}$ when for corresponding functions $\mathbf{f}$ $\ $and $f$ the
condition $\ ^{\mathbf{F(V)}}\mathbf{Af}(\ ^{E}\mathbf{r})=\ ^{\mathbf{P}}%
\mathbf{A}f\mathbf{(u,v).}$ The N--adapted diffusion operator in the phase
space is given by
\begin{equation*}
\ ^{\mathbf{P}}\mathbf{A=}\ ^{v}\triangle +\mathbf{e}_{\ \alpha ^{\prime
}}^{\alpha }(u^{\beta })v^{\alpha ^{\prime }}\mathbf{e}_{\alpha }+B^{%
\widehat{\alpha }}\frac{\partial }{\partial v^{\widehat{\alpha }}},
\end{equation*}%
where the Laplace--Beltrami operator in the hyperbolic velocity space is
\begin{eqnarray*}
\ ^{v}\triangle &=&\delta ^{\widehat{\alpha }^{\prime }\widehat{\beta }%
^{\prime }}E_{\widehat{\alpha }^{\prime }}^{\widehat{\alpha }}\frac{\partial
}{\partial v^{\widehat{\alpha }}}E_{\widehat{\beta }^{\prime }}^{\widehat{%
\beta }}\frac{\partial }{\partial v^{\widehat{\beta }}} \\
&=&h^{\widehat{\alpha }\widehat{\beta }}\left( \frac{\partial ^{2}}{\partial
v^{\widehat{\alpha }}\partial v^{\widehat{\beta }}}+\gamma _{\widehat{\alpha
}\widehat{\beta }}^{\widehat{\varepsilon }}\frac{\partial }{\partial v^{%
\widehat{\varepsilon }}}\right) =\frac{1}{\sqrt{|h_{\widehat{\alpha }%
\widehat{\beta }}|}}\frac{\partial }{\partial v^{\widehat{\varepsilon }}}%
\left( \sqrt{|h_{\widehat{\alpha }\widehat{\beta }}|}h^{\widehat{\varepsilon
}\widehat{\mu }}\frac{\partial }{\partial v^{\widehat{\mu }}}\right).
\end{eqnarray*}
This operator is self--adjoint, $\ ^{v}\triangle =\left( \ ^{v}\triangle
\right) ^{+}.$

The corresponding backward Kolmogorov equation for the N--adapted general
relativistic stochastic processes (\ref{fpdiffg}) is written in the form
\begin{equation*}
\frac{\delta }{\partial \tau }\varphi (\tau ,\mathbf{u,v})=\ ^{\mathbf{P}}%
\mathbf{A}\varphi (\tau ,\mathbf{u,v}).
\end{equation*}%
Using the adjoint of d--operator$\ ^{\mathbf{P}}\mathbf{A,}$ it is possible
to construct the corresponding Fokker--Planck equation in phase space (this
equation in general relativity is also called the Kramer equation). \
Introducing the probability density function $\Phi :=\varphi (\tau ,\mathbf{%
u,v})$ (as the transition probability $\Phi (\mathbf{u,v},\tau |\mathbf{u}%
_{0},$ $\mathbf{v}_{0},\tau =0)$), we write the Fokker--Planck N--adapted
equation in general relativity on a nonholonomic spacetime with d--metric $%
\mathbf{g}_{\alpha \beta }$ (\ref{dm}) and 3--d hyperbolic metric $h_{%
\widehat{\alpha }\widehat{\beta }}$:%
\begin{equation}
\frac{\delta \Phi }{\partial \tau } =-\frac{v^{\alpha ^{\prime }}}{\sqrt{|%
\mathbf{g}_{\alpha \beta }|}}\mathbf{e}_{\gamma }\left( \sqrt{|\mathbf{g}%
_{\alpha \beta }|}\mathbf{e}_{\ \alpha ^{\prime }}^{\gamma }(u^{\beta })\Phi
\right)  -\frac{1}{\sqrt{|h_{\widehat{\alpha }\widehat{\beta }}|}}\frac{\partial }{%
\partial v^{\widehat{\varepsilon }}}\left( \sqrt{|h_{\widehat{\alpha }%
\widehat{\beta }}|}B^{\widehat{\alpha }}\Phi \right) +\frac{\rho }{2}\
^{v}\triangle \Phi ,   \label{reldifeq}
\end{equation}%
where the first two terms in the right side are the divergence d--operators
in, respectively, the position and velocity spaces.

We use the phase--space proper time $\tau $ in the N--adapted general
diffusion equation (\ref{fpdiffg}) (see detailed explanations in \cite{ikeda}
how nontrivial torsion terms can be included in the drift coefficients with
additional terms in $B^{\widehat{\alpha }}$). Alternatively, we can consider
parametrizations in terms of coordinate time (in this work, $u^{3}=t$)\
which is \ convenient for introducing gravitational and external forces
fields. In such cases, the observer time with N--adapted infinitesimal
element $\delta u^{3}=\delta y^{3}=\mathbf{e}_{\ \alpha ^{\prime
}}^{3}(u^{\beta })v^{\alpha ^{\prime }}\delta \tau $ is a function of the
proper time $\tau $ and the space and velocity variables. A simple analogy
with non--relativistic diffusion formula is possible for such frames of
references when $\mathbf{e}_{\ \widehat{\alpha }^{\prime }}^{3}=0$ which can
be introduced in N--adapted Arnovitt--Deser--Misner (ADM). We can introduce $\delta u^{3}=\delta y^{3}=%
\sqrt{|h_{33}(u^{\beta })|}v^{3}\delta \tau ,$ where $v^{3}=\left[
1+(v^{1})^{2}+(v^{2})^{2}+(v^{4})^{2}\right] ^{1/2}.$ Such a model of
general relativistic diffusion in ADM variables, for trivial N--splitting is
elaborated in \cite{herrmann1} (see formulas (36)--(46) in that work). To
generate exact stochastic solutions of Einstein equations we have to
consider nontrivial N--connection structures which makes the
stochastic/diffusion theory more complex but allows us to separate and
integrate the fundamental field equations.

\section{Exact Stochastic Solutions in Gravity}

\label{seses}The anholonomic deformation/frame method of constructing exact
solutions in gravity \cite%
{vexsol2,vexsol3,vncg2} can be extended
to a formalism with stochastic processes and diffusion. In this section, we
provide necessary geometric preliminaries on gravitational field equations
on nonholonomic (pseudo) Riemannian manifolds, show how such equations can
be formally integrated in very general forms and provide some general
criteria/conditions when certain components of metrics and connections are
induced by stochastic generating functions and corresponding diffusion
processes.

\subsection{The Einstein equations on nonholonomic manifolds}

In standard form, the Einstein equations on a nonholonomic (pseudo)
Riemannian manifold (spacetime) $\mathbf{V}$ are written in terms of the
Ricci tensor, $R_{\ \beta \delta },$ and scalar curvature, $R,$ for the
Levi--Civita connection $\nabla ,$ for a given source, i.e. energy--momentum
tensor for matter, $T_{\alpha \beta },$\footnote{%
In brief, \ the gravitational field equations in general relativity are
defined geometrically in this form: Denoting by $\nabla =\{\Gamma _{\ \beta
\gamma }^{\alpha }\}$ the Levi--Civita connection (uniquely defined by a
given tensor $\mathbf{g}$ to be metric compatible, $\nabla \mathbf{g}=0,$
and with zero torsion), the coefficients of necessary tensors are computed
with respect to an arbitrary local frame basis $e_{\alpha }=(e_{i},e_{a})$
and its dual basis $e^{\beta }=(e^{j},e^{b}).$ Using the Riemannian
curvature tensor $\mathcal{R}=\{R_{\ \beta \gamma \delta }^{\alpha }\}$ of $%
\nabla ,$ we define the Ricci tensor, $\mathcal{R}ic=\{R_{\ \beta \delta
}\doteqdot R_{\ \beta \alpha \delta }^{\alpha }\},$ compute the scalar
curvature $R\doteqdot g^{\beta \delta }R_{\ \beta \delta },$ where $g^{\beta
\delta }$ is inverse to $g_{\alpha \beta }.$}
\begin{equation}
R_{\ \beta \delta }-\frac{1}{2}g_{\beta \delta }R=\varkappa T_{\beta \delta
},  \label{einsteq}
\end{equation}%
where $\varkappa =const.$ It is not possible to integrate analytically, in
general form, this system of partial differential equations because of it
generic nonlinearity and complexity. \ In the above mentioned works (see
also references therein), we proved in details that very general integral
varieties can be constructed \ if we rewrite the equations (\ref{einsteq})
in terms of, for instance, the canonical d--connection $\widehat{\mathbf{D}}%
, $
\begin{eqnarray}
&&\widehat{\mathbf{R}}_{\ \beta \delta }-\frac{1}{2}\mathbf{g}_{\beta \delta
}\ ^{s}R=\mathbf{\Upsilon }_{\beta \delta },  \label{cdeinst} \\
&&\widehat{L}_{aj}^{c}=e_{a}(N_{j}^{c}),\ \widehat{C}_{jb}^{i}=0,\ \Omega
_{\ ji}^{a}=0.  \label{lcconstr}
\end{eqnarray}%
In the above formulas, $\widehat{\mathbf{R}}_{\ \beta \delta }$ is the Ricci
tensor for $\widehat{\mathbf{\Gamma }}_{\ \alpha \beta }^{\gamma },\ ^{s}R=%
\mathbf{g}^{\beta \delta }\widehat{\mathbf{R}}_{\ \beta \delta }$ and $%
\mathbf{\Upsilon }_{\beta \delta }$ is constructed for the same metric but
with $\ \widehat{\mathbf{D}},$ similarly to formulas with $\nabla ,$ when $%
\mathbf{\Upsilon }_{\beta \delta }\rightarrow \varkappa T_{\beta \delta }$ \
for $\widehat{\mathbf{D}}\rightarrow \nabla .$ If the constraints (\ref%
{lcconstr}) are satisfied the tensors $\widehat{\mathbf{T}}_{\ \alpha \beta
}^{\gamma }$ (\ref{dtors}) and $Z_{\ \alpha \beta }^{\gamma }$ from (\ref%
{deflc}) are zero and (\ref{cdeinst}) are equivalent to (\ref{einsteq}).
Using nonholonomic deformations and ''non--tensor'' transformation laws for
the coefficients of the linear and d--connections, we can satisfy the
condition $\widehat{\mathbf{\Gamma }}_{\ \alpha \beta }^{\gamma }=\Gamma _{\
\alpha \beta }^{\gamma },$ with respect to N--adapted frames (\ref{ddr}) and
(\ref{ddf}), see (\ref{deflc}), even $\widehat{\mathbf{D}}\neq \nabla .$

Any exact solution of Einstein equations can be parametrized as a metric (%
\ref{gsol}). In classical gravity the coefficients $g_{ij}=diag[g_{i}=\eta
_{i}\ ^{\circ }g_{i}]$ , $h_{ab}=diag[h_{a}=\eta _{a}\ ^{\circ }h_{a}]$ and $%
N_{k}^{3}=w_{i}=\eta _{i}^{3}\ ^{\circ }w_{i},$ $N_{k}^{4}=n_{i}=\eta
_{i}^{4}\ ^{\circ }n_{i}$ are certain smooth and/or singular ''non--random''
classes of real functions defining, for instance, a black hole/worm hole
/cosmological etc solution. The main goal of this paper is to prove that we
can construct new classes of Einstein equations, generated by some
stochastic/random gravitational 'polarizations' $\eta _{\alpha }$ and $\eta
_{i}^{a}$ \ when the nonholonomic deformation of metric $\ ^{\circ }\mathbf{g%
}\mathbf{=}[\ ^{\circ }g_{i},\ ^{\circ }h_{a},\ ^{\circ
}N_{k}^{a}]\rightarrow \ ^{\eta }\mathbf{g}\mathbf{=}[\
g_{i},h_{a},N_{k}^{a}]$ result both in solutions of equations (\ref{cdeinst}%
) and (\ref{lcconstr}), or (\ref{einsteq}), and any type of
stochastic/diffusion equation (for various purposes, we can consider certain
variants of relativistic diffusion, backward Kolmogorov, Fokker--Planck etc
equations). For explicit constructions, we can consider that $\ ^{\circ }%
\mathbf{g}$ is a (pseudo) Riemannian metric (which can be, or not, a
solution of classical Einstein equations) but impose the condition that a
metric $\ ^{\eta }\mathbf{g}$ define a solution (or a class of solutions) of
(\ref{cdeinst}) when certain coefficients are additionally generated by some
stochastic/diffusion processes in curved spaces.

Various types of stochastic generalizations of Einstein manifolds (with
mixed types of non--random and random processes, mutual diffusion of
gravitational and matter fields etc) are called in this work as stochastic
Einstein spaces. Here we emphasize that because of generic nonlinearity of
gravitational field equations the solutions may be with chaos, stochastics,
fractional behavior etc even we may put certain well defined classical
boundary/initial conditions on integration functions. Such nonlinear
stochastic gravitational and matter field configurations are with a very
complex and mixed random--sure spacetime structure and very sophisticate
rules of stochastic and nonholonomic differentiation and integration.

\subsection{Generating stochastic solutions of Einstein equations}

For an ansatz of type (\ref{gsol}), the Einstein equations (\ref{cdeinst})
for $\widehat{\mathbf{D}}$ with a general source of type\footnote{%
parametrization of energy--momentum tensors in the above presented form are
possible by corresponding nonholonomic frame and/or coordinate frame
transform for various types of matter sources, including some general and
important cases with cosmological constants and various models of locally
anisotropic fluid/scalar field/ spinor/ gauge fields interactions on curved
spaces} $\Upsilon _{\ \ \beta }^{\alpha }=diag[\Upsilon _{\gamma };\Upsilon
_{1}=\Upsilon _{2}=\Upsilon _{2}(x^{k},v);\Upsilon _{3}=\Upsilon
_{4}=\Upsilon _{4}(x^{k})]$ transform into a system of nonlinear partial
differential equations with separation of equations for h-- and
v--components of metric and N--connection coefficients, {\small
\begin{eqnarray}
\widehat{R}_{1}^{1} &=&\widehat{R}_{2}^{2}=\frac{-1}{2g_{1}g_{2}}%
[g_{2}^{\bullet \bullet }-\frac{g_{1}^{\bullet }g_{2}^{\bullet }}{2g_{1}}-%
\frac{\left( g_{2}^{\bullet }\right) ^{2}}{2g_{2}}+g_{1}^{\prime \prime }-%
\frac{g_{1}^{\prime }g_{2}^{\prime }}{2g_{2}}-\frac{(g_{1}^{\prime })^{2}}{%
2g_{1}}]=-\Upsilon _{4}(x^{k}),  \label{eq1} \\
\widehat{R}_{3}^{3} &=&\widehat{R}_{4}^{4}=-\frac{1}{2h_{3}h_{4}}%
[h_{4}^{\ast \ast }-\frac{\left( h_{4}^{\ast }\right) ^{2}}{2h_{4}}-\frac{%
h_{3}^{\ast }h_{4}^{\ast }}{2h_{3}}]=-\Upsilon _{2}(x^{k},v),  \label{eq2} \\
\widehat{R}_{3k} &=&\frac{w_{k}}{2h_{4}}[h_{4}^{\ast \ast }-\frac{\left(
h_{4}^{\ast }\right) ^{2}}{2h_{4}}-\frac{h_{3}^{\ast }h_{4}^{\ast }}{2h_{3}}%
]+\frac{h_{4}^{\ast }}{4h_{4}}\left( \frac{\partial _{k}h_{3}}{h_{3}}+\frac{%
\partial _{k}h_{4}}{h_{4}}\right) -\frac{\partial _{k}h_{4}^{\ast }}{2h_{4}}%
=0,  \label{eq3} \\
\widehat{R}_{4k} &=&\frac{h_{4}}{2h_{3}}n_{k}^{\ast \ast }+\left( \frac{h_{4}%
}{h_{3}}h_{3}^{\ast }-\frac{3}{2}h_{4}^{\ast }\right) \frac{n_{k}^{\ast }}{%
2h_{3}}=0,  \label{eq4} \\
w_{i}^{\ast } &=&\mathbf{e}_{i}\ln |h_{4}|,\mathbf{e}_{k}w_{i}=\mathbf{e}%
_{i}w_{k},\ n_{i}^{\ast }=0,\ \partial _{i}n_{k}=\partial _{k}n_{i}
\label{lcconstr1}
\end{eqnarray}
}In brief, we wrote the partial derivatives in the form $a^{\bullet
}=\partial a/\partial x^{1},$\ $a^{\prime }=\partial a/\partial x^{2},$\ $%
a^{\ast }=\partial a/\partial v.$ The ansatz (\ref{gsol}) and resulting
system of equations does not depend on variable $y^{4}$ (we do not have
terms with $\partial /\partial y^{4},$ i.e. our original ansatz was taken
with one Killing symmetry; see references \cite{vexsol2,vexsol3} how to
construct exact solutions with general ''non--Killing'' symmetries). The
constraints (\ref{lcconstr1}) have to be imposed additionally if we wont to
satisfy the conditions (\ref{lcconstr}) and generate solutions in general
relativity just for the Levi--Civita connection $\nabla .$

The above system of equations can be integrated in very general forms \cite%
{vexsol2,vexsol3}, in sure variables, by integrating step by step
the equations \ beginning with (\ref{eq1}), which independent from another
ones (relating two coefficients of h--metric, $g_{1}$ and $g_{2},$ and
source $\Upsilon _{4},$ all depending on two variables $x^{k}$), then (\ref%
{eq2}) (relating two coefficients of v--metric, $h_{3}$ and $h_{4},$ and
source $\Upsilon _{2}$, all depending on three variables $x^{k},v,$ where we
take $v=t;$ the time coordinate can be related for ansatz of type (\ref{gsol}%
) with the parameter $\tau $ via relation $\delta \tau =\sqrt{|h_{3}|}\delta
t$). The equations (\ref{eq3}) is an algebraic one for N--connection
coefficients $w_{k}$ and we have to integrate two times the equations (\ref%
{eq4}) \ in order to compute the N--connection coefficients $n_{k};$ in both
cases, we can generate sure or stochastic solutions, depending on the type
of solutions we have for $h_{3}$ and/or $h_{4}.$ Finally, the equations (\ref%
{lcconstr1}) \ impose additional constraints on integral varieties for (\ref%
{eq1})--(\ref{eq4}) eliminating induced (by off--diagonal coefficients of
metric, i.e. N--connection) torsion and constraining the system of equations
and solutions to be just for $\nabla .$

The h-- and v--separation allows us not only to generalize the constructions
for stochastic metrics and N--connections but also to analyze the conditions
when a non--random h--metric structure model gravitational diffusion
processes in the v--subspace.

\subsection{Stochastic solutions with $h_{3,4}^{\ast }\neq 0$ and $\Upsilon
_{2,4}\neq 0$}

We consider a metric (\ref{gsol}) when
\begin{eqnarray}
\ ^{\eta }\mathbf{g} &\mathbf{=}&e^{\psi (x^{k})}{dx^{i}\otimes dx^{i}}%
+h_{3}(x^{k},t)\mathbf{e}^{3}{\otimes }\mathbf{e}^{3}+h_{4}(x^{k},t)\mathbf{e%
}^{4}{\otimes }\mathbf{e}^{4},  \label{genans} \\
\mathbf{e}^{3} &=&dt+w_{i}(x^{k},t)dx^{i},\mathbf{e}%
^{4}=dy^{4}+n_{i}(x^{k},t)dx^{i}  \notag
\end{eqnarray}%
is supposed to be a solution of (\ref{eq1})--(\ref{eq4}) with $g_{1}=$ $%
g_{2}=e^{\psi (x^{k})}$ defining a non--random h--metric, as a solution of
2-d Laplace equation (the h--components of the Einstein equations transform
to this simple equation which can be used for generating sure, or random
solutions)%
\begin{equation}
\ddot{\psi}+\psi ^{\prime \prime }=2\Upsilon _{4}(x^{k}),  \label{4ep1a}
\end{equation}%
and $h_{3},h_{4}$ defining a stochastic v--metric (in result, $w_{i}$ and/or
$n_{i}$ can be also stochastic solutions).

Introducing values
\begin{equation}
~\phi =\ln |\frac{h_{4}^{\ast }}{\sqrt{|h_{3}h_{4}|}}|,\ \alpha
_{i}=h_{4}^{\ast }\partial _{i}\phi ,\ \beta =h_{4}^{\ast }\ \phi ^{\ast },\
\gamma =\left( \ln |h_{4}|^{3/2}/|h_{3}|\right) ^{\ast },  \label{auxphi}
\end{equation}%
the equations (\ref{eq3}),(\ref{eq4}) are respectively written in the form%
\begin{eqnarray}
\beta w_{i}+\alpha _{i} &=&0,  \label{4ep3a} \\
n_{i}^{\ast \ast }+\gamma n_{i}^{\ast } &=&0  \label{4ep4a}
\end{eqnarray}%
The type of solutions for N--connection coefficients depend explicitly on
the type of solutions we construct/choose for the v--metric.

\subsubsection{Non--random/sure solutions}

For sure coefficients, the equation (\ref{eq2}) transform into
\begin{equation}
h_{4}^{\ast }=2h_{3}h_{4}\Upsilon _{2}(x^{i},t)/\phi ^{\ast }.  \label{4ep2a}
\end{equation}%
If $h_{4}^{\ast }\neq 0;\Upsilon _{2}\neq 0,$ we get $\phi ^{\ast }\neq 0.$
Prescribing any non--constant $\phi =\phi (x^{i},t)$ as a generating
function, we can construct exact solutions of (\ref{4ep2a})--(\ref{4ep4a}).
Integrating on $t,$ in order to determine $h_{3},$ $h_{4}$ and $n_{i},$ and
solving algebraic equations, for $w_{i},$ we get
\begin{eqnarray}
h_{3} &=&\pm \ \frac{|\phi ^{\ast }(x^{i},t)|}{\Upsilon _{2}},\ h_{4}=\
^{0}h_{4}(x^{k})\pm \ 2\int \frac{(\exp [2\ \phi (x^{k},t)])^{\ast }}{%
\Upsilon _{2}}dt,\   \label{gsol1} \\
w_{i} &=&-\partial _{i}\phi /\phi ^{\ast },\ n_{i}=\ ^{1}n_{k}\left(
x^{i}\right) +\ ^{2}n_{k}\left( x^{i}\right) \int [h_{3}/(\sqrt{|h_{4}|}%
)^{3}]dt,  \notag
\end{eqnarray}%
where $\ ^{0}h_{4}(x^{k}),\ ^{1}n_{k}\left( x^{i}\right) $ and $\
^{2}n_{k}\left( x^{i}\right) $ are integration functions. We have to fix a
corresponding sign $\pm $ in order to generate a necessary local signature
of type $(++-+)$ for some chosen $\phi ,\Upsilon _{2}$ and $\Upsilon _{4}.$
Such solutions include as particular cases the classes of solutions for a
nontrivial cosmological constant $\Upsilon _{i}=\lambda ,$ or nonholonomic
configurations with polarizations of such constants, $\lambda \rightarrow \
^{h}\lambda (x^{k})=\Upsilon _{4}(x^{k})$ and $\lambda \rightarrow \
^{v}\lambda (x^{k},t)=\Upsilon _{2}(x^{k},t).$

In Refs. \cite{vexsol2,vexsol3,vncg2}, we studied various
classes of parametric and non--parametric exact solutions with coefficients
of type (\ref{4ep1a}) and (\ref{4ep2a}) describing black ellipsoids, locally
anisotropic wormholes and cosmological solutions and their noncommutative
generalizations. \ Those \ constructions can be generalized for stochastic
Einstein spaces if \ $\phi $ and/or $\Upsilon _{2}$ are taken to be \
certain stochastic functions.

\subsubsection{Metrics with h--diffusion of generating functions}

This class of solutions if we chose, for instance, a generating function
\begin{equation}
\phi (x^{k},t)\rightarrow \underline{\phi }(x^{k},t)=\phi (x^{k},t)+\varpi
\widetilde{\phi }(x^{k},t),  \label{gsol1a}
\end{equation}%
where $\widetilde{\phi }(x^{k},t)\sim \underline{f}(\tau ,\mathbf{u})=%
\underline{f}(\tau ,x^{i})$ is defined as as diffusion process from h--space
to v--space by a corresponding backward Kolmogorov equation for the
N--adapted general relativistic stochastic processes (\ref{kfpnh}) and/or (%
\ref{fpnh}) (in order to state the solutions in \ exact form, we consider
only h--operators and fix $\mathbf{A}^{\beta }=0$ in $\ ^{\mathbf{V}}%
\widetilde{\mathbf{A}}$ (\ref{diffusd})), where $\tau \rightarrow t,$ $%
\mathbf{v=v}_{0}=const$ and the Laplace--Beltrami d--operator $\widehat{%
\bigtriangleup }$ (\ref{dlb}) is computed as $\widehat{\bigtriangleup }=h%
\widehat{\bigtriangleup }$ for $g_{ij}=\delta _{ij}e^{\psi (x^{k})}.$ We can
consider $\varpi $ as a real parameter which for $\varpi =0$ transforms our
system into a non--random one. For solutions with one Killing symmetry such
parameters can be always introduced.
Nevertheless, there are substantial differences between the parameter $%
\varpi $ and those for sure solutions with superposition of Killing
symmetries and/or with noncommutative parameters $\theta ,$ see \cite%
{vncg2}. In this work, a nontrivial $\varpi $ results in
stochastic behavior of (some) coefficients of metrics and connections, i.e.
of gravitational fields.

The generalized Kolmogorov backward equation (describing diffusion of
gravitational h--components on $\tau ,$ i. e. into $v$--components) is
\begin{eqnarray}
\partial _{\tau }\underline{f}(\tau ,x^{i}) &=&\ h\widehat{\bigtriangleup }%
\underline{f}((\tau ,x^{i})=h\widehat{\bigtriangleup }\underline{f}((\tau
,x^{i}),  \label{kfpnh1} \\
\underline{f}(0,x^{i}) &=&f(x^{i}).  \notag
\end{eqnarray}%
The corresponding generalized Fokker--Planck equation is
\begin{equation}
\partial _{\tau }\digamma =\frac{\rho }{2}h\widehat{\bigtriangleup }\digamma
,  \label{fpnh1}
\end{equation}%
where $\digamma =\digamma (\tau ,\ ^{1}x^{i}\mathbf{;}0\mathbf{,\ }%
^{2}x^{i}) $ is the transition probability with the initial condition $%
\digamma (0,\ ^{1}x^{i}\mathbf{;}0\mathbf{,\ }^{2}x^{i})=\delta (\ ^{1}x^{i}%
\mathbf{-\ }^{2}x^{i})$ for any two points $\ ^{1}x^{i}\mathbf{,\ }^{2}x^{i}%
\mathbf{\in } $\ $\mathbf{V}$ and adequate \ boundary conditions at
infinity. The same equations can be considered for the probability density $%
\varphi (\tau ,x^{i})$ but for the initial condition $\varphi (\tau
=0,x^{i})=\ ^{0}\varphi (x^{i}).$ \ Stochastic components $\widetilde{\phi }%
(x^{k},t)$ in (\ref{gsol1}) \ and (\ref{gsol1a}) generated as diffusion
solutions of (\ref{kfpnh1}) and/or (\ref{fpnh1}) provide some ''simple'' and
explicit examples of stochastic metrics in Einstein gravity, all
distinguished by corresponding nonholonomic constraints. Under general
frame/coordinate transforms, the parametizatons change into mixed
holonomic/nonholonomic and non--random/stochastic variables both on h-- and
v--subspaces. More general types of nonholonomic diffusion equations can be
also considered but the solutions will be generated in non--explicit forms.

Fixing $\ ^{2}n_{k}\left( x^{i}\right) =0$ in (\ref{gsol1}), we generate
non--random solutions for the N--connection components $n_{i}.$ But $w_{i}$
will have positively a stochastic character, as well both $h_{3}$ and $%
h_{4}, $ if the generating function $\underline{\phi }(x^{k},t)$ contains a
nontrivial diffusion component. Such random gravitational processes can be
induced even as vacuum gravitational configurations for smooth coefficients
of Einstein equations

\subsubsection{Solutions induced by random sources}

There is an alternative possibility to generate random gravitational
processes than those with stochastic generating functions. This can be seen
from formulas (\ref{4ep2a}) and (\ref{gsol1}) for certain random/noise
behavior of source $\Upsilon _{2}.$ Such constructions are similar to those
for \textit{stochastic gravity} (based on the Einstein--Langeven equation
with additional sources due to the noise kernel, see a review and references
in \cite{hu}). Nevertheless, it should be mentioned here that our
anholonomic deformation method is different from the methods of deriving
solutions in stochastic gravity. Via nonholonomic transforms/deformations we
can generate in a "non--perturbative" manner different classes of exact
solutions with generic off--diagonal terms and mixed holonomic--nonholonomic
variables and matter sources.

In a particular case, we can consider that a stochastic $\Upsilon _{2}$ can
be constructed as a random vacuum polarization of gravitational constant,
when vacuum gravitational \ fields are nonholonomically distorted as a
diffusion process, for instance, modelled by solutions of equations (\ref%
{kfpnh1}) and/or (\ref{fpnh1}) associated to some effective matter
fiels/cosmological constant.

\subsubsection{Diffusion to the Levi--Civita conditions}

In order to construct exact solutions for the Levi--Civita connection, i.e.
of standard form of Einstein equations written with respect to N--adapted
frames, we have to constrain the coefficients (\ref{gsol1}) of metric (\ref%
{genans}) to satisfy the conditions (\ref{lcconstr1}). There are various
possibilities. For instance, we may consider a classical sure metric (which
may be, or not, a solution of gravitational field equations with nontrivial
distortion) and constrain the system in such a way that gravitational
diffusion will result in stochastic vacuum or Einstein spaces. There are
scenarios when a sure Einstein configuration is stochastically transformed
into nonholonomic configuration with nontrivial distortion of of the
Levi--Civita connection.

The above mentioned gravitational diffusion evolution models \ depend on the
class of additional constraints we impose on generating and integration
functions. For instance, we can chose that $\ ^{2}n_{k}\left( x^{i}\right)
=0 $ and $\ ^{1}n_{k}\left( x^{i}\right) $ are any functions satisfying the
non--random conditions $\ \partial _{i}\ ^{1}n_{k}=\partial _{k}\ ^{1}n_{i}.$
The constraints for $\phi (x^{k},t)$ can be for random, sure and/or of mixed
nature variables, following from constraints on N--connection coefficients $%
w_{i}=-\partial _{i}\phi /\phi ^{\ast },$
\begin{eqnarray}
\left( w_{i}[\phi ]\right) ^{\ast }+w_{i}[\phi ]\left( h_{4}[\phi ]\right)
^{\ast }+\partial _{i}h_{4}[\phi ]=0, &&  \notag \\
\partial _{i}\ w_{k}[\phi ]=\partial _{k}\ w_{i}[\phi ], &&  \label{auxc1}
\end{eqnarray}%
where, for instance, we denoted by $h_{4}[\phi ]$ the functional dependence
on $\phi .$ So, if $\phi $ is stochastic, we have to consider that (\ref%
{auxc1}) are some relations on mathematical expectations etc. Such
conditions are always satisfied for cosmological solutions with $\phi =\phi
(t)$ or if $\phi =const$ (in the last case $w_{i}(x^{k},t)$ can be any
non--random and/or stochastic functions as follows from (\ref{4ep3a}) with
zero $\beta $ and $\alpha _{i},$ see (\ref{auxphi})).

\subsection{Special cases of stochastic Einstein spaces}

We can construct such solutions for certain special parametrizations of
coefficients for ansatz (\ref{gsol1}) subjected to the condition to be
solutions of equations (\ref{4ep1a})--(\ref{4ep4a}) and certain nonholonomic
diffusion equations on curved spaces.

\subsubsection{Vacuum gravitational diffusion with $h_{4}^{\ast }=0$}

The equation (\ref{eq2}) can be solved for such a case, $h_{4}^{\ast }=0,$
only if $\Upsilon _{2}=0.$ So, in N--adapted frames, the v--components of
gravitational equations are vacuum ones. We can consider any functions $%
w_{i}(x^{k},t),$ being sure or random ones, as solutions of (\ref{eq3}), and
its equivalent (\ref{4ep3a}), because the coefficients $\beta $ and $\alpha
_{i},$ see (\ref{auxphi}), are zero.

We find nontrivial values of $n_{i}$ by integrating (\ref{4ep4a}) for $%
h_{4}^{\ast }=0$ and any given $h_{3}$ which results in $n_{i}=\
^{1}n_{k}\left( x^{i}\right) +\ ^{2}n_{k}\left( x^{i}\right) \int h_{3}dt.$
Choosing $h_{3}$ to be a stochastic function, we have to compute $\int
h_{3}dt $ as a Stratonovich stochastic integral, when $\int
h_{3}dt\rightarrow \int \widetilde{h}_{3}\circ d\tau .$ We can consider any $%
g_{1}=g_{2}=e^{\psi (x^{k})},$ with $\psi (x^{k})$ determined by (\ref{4ep1a}%
) for a given $\Upsilon _{4}(x^{k});$ for simplicity, we can consider only
non--random solutions for the h--metric. \

Summarizing the constructions, we get a class of stochastic gravitational
solutions defined by ansatz
\begin{eqnarray}
\ ^{\eta }\mathbf{g} &\mathbf{=}&e^{\psi (x^{k})}{dx^{i}\otimes dx^{i}}%
+h_{3}(x^{k},t)\mathbf{e}^{3}{\otimes }\mathbf{e}^{3}+\ ^{0}h_{4}(x^{k})%
\mathbf{e}^{4}{\otimes }\mathbf{e}^{4},  \label{genans1} \\
\mathbf{e}^{3} &=&dt+w_{i}(x^{k},t)dx^{i},  \notag \\
\mathbf{e}^{4} &=&dy^{4}+[\ ^{1}n_{k}\left( x^{i}\right) +\ ^{2}n_{k}\left(
x^{i}\right) \int h_{3}dt\rightarrow \int \widetilde{h}_{3}\circ d\tau
]dx^{k},  \notag
\end{eqnarray}%
for arbitrary generating stochastic and/or sure functions $%
h_{3}(x^{k},t),w_{i}(x^{k},t),$ $\ ^{0}h_{4}(x^{k})$ and integration
functions $\ ^{1}n_{k}\left( x^{i}\right) $ and $\ ^{2}n_{k}\left(
x^{i}\right) .$

The conditions (\ref{lcconstr1}) selecting from (\ref{genans1}) \ a subclass
of solutions for the Levi--Civita connection transform into the equations
\begin{eqnarray*}
\ ^{2}n_{k}\left( x^{i}\right) =0\ &\mbox{ and }&\partial _{i}\
^{1}n_{k}=\partial _{k}\ ^{1}n_{i}, \\
w_{i}^{\ast }+\partial _{i}\ ^{0}h_{4}=0 &\mbox{ and }&\partial _{i}\
w_{k}=\partial _{k}\ w_{i},
\end{eqnarray*}%
for any such $w_{i}(x^{k},t)$ and $\ ^{0}h_{4}(x^{k}).$ This class of
constraints do not involve the generating function $h_{3}(x^{k},t).$ So, we
can consider sure constraints to the Levi--Civita configurations for a
stochastic $h_{3}.$ In general, we can model nontrivial Levi--Civita
diffusion processes for sure and/or stochastic $h_{3},$ but constraining the
gravitational diffusion process to conditions $h_{4}^{\ast }=0$ and $%
\Upsilon _{2}=0.$

\subsubsection{Gravitational diffusion with $h_{3}^{\ast }=0$ and $%
h_{4}^{\ast }\neq 0$}

Such stochastic spacetimes are defined by ansatz of type
\begin{eqnarray}
\ ^{\eta }\mathbf{g} &\mathbf{=}&e^{\psi (x^{k})}{dx^{i}\otimes dx^{i}}-\
^{0}h_{3}(x^{k})\mathbf{e}^{3}{\otimes }\mathbf{e}^{3}+h_{4}(x^{k},t)\mathbf{%
e}^{4}{\otimes }\mathbf{e}^{4},  \notag \\
\mathbf{e}^{3} &=&dt+w_{i}(x^{k},t)dx^{i},\mathbf{e}%
^{4}=dy^{4}+n_{i}(x^{k},t)dx^{i},  \label{genans2}
\end{eqnarray}%
where $g_{1}=g_{2}=e^{\psi (x^{k})},$ with $\psi (x^{k})$ being a solution
of (\ref{4ep1a}) for any given $\Upsilon _{4}(x^{k}).$ The function $%
h_{4}(x^{k},t)$ must satisfy the equation (\ref{4ep2a}) which for $%
h_{3}^{\ast }=0$ is just
\begin{equation*}
h_{4}^{\ast \ast }-\frac{\left( h_{4}^{\ast }\right) ^{2}}{2h_{4}}-2\
^{0}h_{3}h_{4}\Upsilon _{2}(x^{k},t)=0.
\end{equation*}%
The N--connection coefficients are
\begin{equation*}
w_{i}=-\partial _{i}\widetilde{\phi }/\widetilde{\phi }^{\ast },~n_{i}=\
^{1}n_{k}\left( x^{i}\right) +\ ^{2}n_{k}\left( x^{i}\right) \int [1/(\sqrt{%
|h_{4}|})^{3}]dt,
\end{equation*}%
when $\widetilde{\phi }=\ln |h_{4}^{\ast }/\sqrt{|\ ^{0}h_{3}h_{4}|}|.$
Fixing $\ ^{0}h_{3}=0,$ we can always eliminate possible stochastic
contributions from $\Upsilon _{2}$ with a sure value $h_{4}$ being a
solution of $h_{4}^{\ast \ast }=\left( h_{4}^{\ast }\right) ^{2}/2h_{4}.$
This impose us to consider classical non--random values for $\widetilde{\phi
}$ and, as consequences, for $w_{i}$ and $n_{i}.$

If we consider any $\ ^{0}h_{3}\neq 0,$ the stochastic gravitational
processes will be induced by ''noise'' in $\Upsilon _{2}.$ So, the
stochastic metrics of type (\ref{genans2}) are generically defined by a
stochastic source $\Upsilon _{2}.$ Such nonholonomic stochastic
gravitational configurations are very different by those described by ansatz
(\ref{genans1}) when the gravitational diffusion is of generic vacuum type.

The Levi--Civita configurations for (\ref{genans2}) are selected by the
conditions (\ref{lcconstr1}) which, for this case, are satisfied if
\begin{eqnarray*}
\ ^{2}n_{k}\left( x^{i}\right) =0\ \mbox{ and }\partial _{i}\
^{1}n_{k}=\partial _{k}\ ^{1}n_{i},&& \\
\mbox{ and \ } \left( w_{i}[\widetilde{\phi }]\right) ^{\ast }+w_{i}[%
\widetilde{\phi }]\left( h_{4}[\widetilde{\phi }]\right) ^{\ast }+\partial
_{i}h_{4}[\widetilde{\phi }]=0, && \\
\partial _{i}\ w_{k}[\widetilde{\phi }]=\partial _{k}\ w_{i}[\widetilde{\phi
}]. &&
\end{eqnarray*}%
Such conditions are similar to (\ref{auxc1}) but for a different relation of
v--coefficients of metric to another type of generating function $\widetilde{%
\phi }$ which can be stochastic only for random values of $\Upsilon _{2}.$
They are always satisfied for cosmological solutions with $\widetilde{\phi }=%
\widetilde{\phi }(t)$ or if $\widetilde{\phi }=const.$ In the last case $%
w_{i}(x^{k},t)$ can be any stochastic or sure functions as follows from (\ref%
{4ep3a}) with zero $\beta $ and $\alpha _{i},$ see (\ref{auxphi}). So, for
some special \ configurations, random values of $w_{k}$ can be induced via
nonholonomic vacuum gravitational diffusion even the source $\Upsilon _{2}$
is constrained to be non--random.

\subsubsection{Solutions with constant generating functions}

Fixing $\phi =\phi _{0}=const$ in (\ref{auxphi}), with $h_{3}^{\ast }\neq 0$
and $h_{4}^{\ast }\neq 0,$ we can express the general solutions of (\ref%
{4ep1a})--(\ref{4ep4a}) in the form
\begin{eqnarray}
\ ^{\eta }\mathbf{g} &\mathbf{=}&e^{\psi (x^{k})}{dx^{i}\otimes dx^{i}}-\
^{0}h^{2}\ \left[ f^{\ast }\left( x^{i},t\right) \right] ^{2}|\varsigma
_{\Upsilon }\left( x^{i},t\right) |\mathbf{e}^{3}{\otimes }\mathbf{e}^{3}
+f^{2}\left( x^{i},t\right) \mathbf{e}^{4}{\otimes }\mathbf{e}^{4},  \notag
\\
\mathbf{e}^{3} &=&dt+w_{i}(x^{k},t)dx^{i},\ \mathbf{e}^{4}=dy^{4}+n_{k}%
\left( x^{i},t\right) dx^{i},  \label{genans3}
\end{eqnarray}%
where $^{0}h=const,$ $g_{1}=g_{2}=e^{\psi (x^{k})},$ with $\psi (x^{k})$
being a solution of (\ref{4ep1a}) for any given $\Upsilon _{4}(x^{k}),$ and
\begin{equation*}
\varsigma _{\Upsilon }\left( x^{i},t\right) =\varsigma _{4[0]}\left(
x^{i}\right) -\frac{h_{0}^{2}}{16}\int \Upsilon _{2}(x^{k},t)[f^{2}\left(
x^{i},t\right) ]^{2}dt.
\end{equation*}%
In a stochastic sense, we should consider $\int ...dt$ $\rightarrow \int
Stratonovich\ \delta \tau .$

The N--connection coefficients $N_{i}^{3}=w_{i}(x^{k},t)\  %
N_{i}^{4}=n_{i}(x^{k},t)$ are
\begin{eqnarray}
w_{i}&=&-\frac{\partial _{i}\varsigma _{\Upsilon }\left( x^{k},t\right) }{%
\varsigma _{\Upsilon }^{\ast }\left( x^{k},t\right) }  \label{gensol1w} \\
n_{k}&=&\ ^{1}n_{k}\left( x^{i}\right) +\ ^{2}n_{k}\left( x^{i}\right) \int
\frac{\left[ f^{\ast }\left( x^{i},t\right) \right] ^{2}}{\left[ f\left(
x^{i},t\right) \right] ^{2}}\varsigma _{\Upsilon }\left( x^{i},t\right) dt,
\label{gensol1n}
\end{eqnarray}%
with necessary generalizations to Stratonovich stochastic integrals.

We must take $\varsigma _{4[0]}\left( x^{i}\right) =\pm 1$ if $\varsigma
_{\Upsilon }\left( x^{i},t\right) =\pm 1$ for $\Upsilon _{2}\rightarrow 0.$
In such a case, the functions $h_{3}=-\ ^{0}h^{2}\ \left[ f^{\ast }\left(
x^{i},t\right) \right] ^{2}$ and $h_{4}=f^{2}\left( x^{i},t\right) $ satisfy
the equation (\ref{4ep2a}) written in the form $\sqrt{|h_{3}|}=\ ^{0}h(\sqrt{%
|h_{4}|})^{\ast },$ which is compatible with the condition $\phi =\phi _{0}.$

The \ gravitational diffusion for this \ class of solutions has two sources:
the first one can be if $\ f\left( x^{i},t\right) $ \ is a random generating
function and the second one is for random sources $\Upsilon _{2}(x^{k},t).$
Gravitational diffusion mix such contributions even we distinguish them
nonholonomically for certain configurations.

The subclass of solutions for the Levi--Civita connection with ansatz of
type (\ref{genans3}) is selected via conditions (\ref{lcconstr1}). We can
chose that $\ ^{2}n_{k}\left( x^{i}\right) =0$ and $\ ^{1}n_{k}\left(
x^{i}\right) $ are any functions satisfying the conditions $\ \partial _{i}\
^{1}n_{k}=\partial _{k}\ ^{1}n_{i}.$ The constraints on values $%
w_{i}=-\partial _{i}\varsigma _{\Upsilon }/\varsigma _{\Upsilon }^{\ast }$
result in constraints on $\varsigma _{\Upsilon },$ which is determined by $%
\Upsilon _{2}$ and $f,$
\begin{eqnarray}
\left( w_{i}[\varsigma _{\Upsilon }]\right) ^{\ast }+w_{i}[\varsigma
_{\Upsilon }]\left( h_{4}[\varsigma _{\Upsilon }]\right) ^{\ast }+\partial
_{i}h_{4}[\varsigma _{\Upsilon }]=0, &&  \notag \\
\partial _{i}\ w_{k}[\varsigma _{\Upsilon }]=\partial _{k}\ w_{i}[\varsigma
_{\Upsilon }], &&  \label{auxc3}
\end{eqnarray}%
where, for instance, we denoted by $h_{4}[\varsigma _{\Upsilon }]$ the
functional dependence on $\varsigma _{\Upsilon }$ including random and
non--random contributions both from vacuum gravitational configurations and
stochastic processes for matter. Such conditions are always satisfied for
cosmological solutions with $f=f(t).$ For $\widehat{\mathbf{D}},$ if $\
\Upsilon _{2}=0$ and $\phi =const,$ the coefficients $w_{i}(x^{k},t)$ can be
arbitrary functions (we can fix $\varsigma _{\Upsilon }=1,$ which does not
impose a functional dependence of $w_{i}$ on $\varsigma _{\Upsilon })$ as
follows from (\ref{4ep3a}) with zero $\beta $ and $\alpha _{i},$ see (\ref%
{auxphi}). \ As in the previous case, such N--connection components can be
stochastic ones as arising from some vacuum gravitational configurations. To
generate solutions for $\nabla $ such $w_{i}$ must be additionally
constrained following formulas (\ref{auxc3}) re--written for $%
w_{i}[\varsigma _{\Upsilon }]\rightarrow w_{i}(x^{k},t)$ and $%
h_{4}[\varsigma _{\Upsilon }]\rightarrow h_{4}\left( x^{i},t\right) .$ Such
scenarios are typical ones with nonholonomic gravitational diffusion from/to
Levi--Civita configurations.

\subsubsection{Non--Killing stochastic solutions}

We note that any sure and/or stochastic solution $\mathbf{g}=\{g_{\alpha
^{\prime }\beta ^{\prime }}(u^{\alpha ^{\prime }})\}$ of the Einstein
equations (\ref{cdeinst}) and/or (\ref{einsteq}) with Killing symmetry $%
\partial /\partial y$ (for local coordinates in the form $y^{3}=t$ and $%
y^{4}=y)$ can be parametrized in a form derived in this section. Using frame
transforms of type $e_{\alpha }=e_{\ \alpha }^{\alpha ^{\prime }}e_{\alpha
^{\prime }},$ with $\mathbf{g}_{\alpha \beta }=e_{\ \alpha }^{\alpha
^{\prime }}e_{\ \beta }^{\beta ^{\prime }}g_{\alpha ^{\prime }\beta ^{\prime
}},$ for any $\mathbf{g}_{\alpha \beta }$ (\ref{gsol}), we relate the class
of such (inhomogeneous) cosmological solutions, for instance, to the family
of metrics of type (\ref{genans}).\footnote{%
We have to solve certain systems of quadratic algebraic equations and define
some $e_{\ \alpha }^{\alpha ^{\prime }}(u^{\beta }),$ choosing a convenient
system of coordinates $u^{\alpha ^{\prime }}=u^{\alpha ^{\prime }}(u^{\beta
}).$} In explicit form, the solutions can be modelled as explicit diffusions
from the h-- to v--components of geometrical/physical objects.

Following our recent results on constructing general solutions in Einstein
gravity and modifications \cite{vexsol2,vexsol3,vncg2}, we
can construct 'non--Killing' solutions depending on all coordinates. Such
general classes of sure solutions can be parametrized in the form
\begin{eqnarray}
\mathbf{g} &\mathbf{=}&+g_{i}(x^{k}){dx^{i}\otimes dx^{i}}+\omega
^{2}(x^{j},t,y)h_{a}(x^{k},t)\mathbf{e}^{a}{\otimes }\mathbf{e}^{a},  \notag
\\
\mathbf{e}^{3} &=&dy^{3}+w_{i}(x^{k},t)dx^{i},\mathbf{e}%
^{4}=dy^{4}+n_{i}(x^{k},t)dx^{i},  \label{ansgensol}
\end{eqnarray}%
for any $\omega $ for which
\begin{equation}
\mathbf{e}_{k}\omega =\partial _{k}\omega +w_{k}\omega ^{\ast
}+n_{k}\partial \omega /\partial y=0,  \label{conf}
\end{equation}
when (\ref{ansgensol}) with $\omega ^{2}=1$ is of type (\ref{gsol}).

A class of stochastic solutions with nontrivial sure factor $\omega $
satisfying the sure conditions (\ref{conf}) with sure values $w_{k} $ and $%
n_{k}$ but stochastic $h_{a}$ can be generated similarly to the case of
nonholonomic sure Killing configurations. In a more general approach, we can
impose the conditions (\ref{conf}) for any stochastic values $\omega
,n_{k},w_{k}$ and $h_{a}.$ With respect to N--adapted frames such conditions
can be satisfied for certain particular, explicit examples, configurations
or in non--explicit forms.

\section{Summary and Conclusions}

\label{sconcl}In this work, we have studied the best strategy, and
elaborated a geometric method, for generating stochastic exact solutions of
Einstein equations in general relativity and modifications.

Our proposal is to follow respectively certain key steps and sub--steps for
constructing deterministic and stochastic solutions for nonholonomic
gravitational and matter fields interactions:

\begin{enumerate}
\item \textbf{Nonholonomic splitting \& formal integration of Einstein eqs}

\begin{itemize}
\item Let us introduce a nonholonomic $2+2$ distribution stating a nonlinear
connection (N--connection) structure $\mathbf{N}:\ T\mathbf{V=}h\mathbf{V}%
\oplus v\mathbf{V}$ with associated N--adapted frames, $\mathbf{e}_{\alpha
}=(\mathbf{e}_{i},\partial _{a})$ and $\mathbf{e}^{\beta }=(e^{i},\mathbf{e}%
^{a}),$ on a (pseudo) Riemannian manifold $\mathbf{V}$ enabled with metric
structure $\mathbf{g,}$ when $\partial _{a}=\partial /\partial y^{a}$ and $%
e^{i}=dx^{i}$ for local coordinates $u^{\beta }=(x^{j},y^{a}),$ or $\mathbf{u%
}=(\mathbf{x,y})$.

\item We adapt all geometric constructions to $\mathbf{N}$ and define the
canonical distinguished connection(d--connection) $\widehat{\mathbf{D}}%
=\nabla +\widehat{\mathbf{Z}},$ where $\nabla $ and $\widehat{\mathbf{Z}}$
are respectively the Levi--Civita connection $\nabla $ and the canonical
distortion distinguished tensor (d--tensor), all defined in a unique metric
compatible form by $\mathbf{g.}$ Any N--connection structure $\mathbf{N=\{}%
N_{i}^{a}(\mathbf{u})\}$ states a conventional horizontal--vertical (h-v)
decompositions of geometric object, for instance, $\widehat{\mathbf{D}}=h%
\widehat{D}\oplus v\widehat{D}$ and $\mathbf{g=}hg\oplus vg,$ $\ $for $%
hg=\{g_{ij}\}$ and $vg=\{h_{ab}\}$ with respect to N--elongated frames, when
$\mathbf{e}_{i}=\partial _{i}-N_{i}^{b}\partial _{b}$ and $\mathbf{e}%
^{a}=dy^{a}+N_{i}^{a}dx^{i}.$

\item We write the Einstein equations (\ref{cdeinst}) and (\ref{lcconstr})
for the geometric data $( \mathbf{g,N,}\widehat{\mathbf{D}});$ the
gravitational filed equations are equivalent to those in standard variables $%
(\mathbf{g,\nabla }),$ see (\ref{einsteq}), for correspondingly defined
energy--momentum and distortion sources.

\item For a general ansatz, we get that the Einstein equations are
equivalent to a system of nonlinear PDE (\ref{eq1})--(\ref{eq4}) and (\ref%
{lcconstr1}), with a ''consequent splitting'' of equations which allows us
to construct general off--diagonal solutions in sure variables; in local
coordinate frames such metrics are parametrized in the form\footnote{%
as a matter of principle, any solution of gravitational field equations with
certain general matter fields sources can be represented in such a form by
corresponding frame and coordinate transforms; we have to involve certain
additional physical considerations, suppositions on symmetry of interactions
and boundary conditions in order to model realistic gravitational
interactions etc.} 
\begin{equation*}
g_{\alpha \beta }(u^{\tau })=q\times \left|
\begin{array}{cccc}
g_{1}+\omega ^{2}(w_{1}^{\ 2}h_{3}+\omega ^{2}(n_{1}^{\ 2}h_{4}) & \omega
^{2}(w_{1}^{\ }w_{2}^{\ }h_{3}+n_{1}^{\ }n_{2}^{\ }h_{4}) & \omega ^{2}\
w_{1}^{\ }h_{3} & \omega ^{2}\ n_{1}^{\ }h_{4} \\
\omega ^{2}(w_{1}^{\ }w_{2}^{\ }h_{3}+n_{1}^{\ }n_{2}^{\ }h_{4}) &
g_{2}+\omega ^{2}(w_{2}^{\ 2}h_{3}+n_{2}^{\ 2}h_{4}) & \omega ^{2}\ w_{2}^{\
}h_{3} & \omega ^{2}\ n_{2}^{\ }h_{4} \\
\omega ^{2}\ w_{1}^{\ }h_{3} & \omega ^{2}\ w_{2}^{\ }h_{3} & h_{3} & 0 \\
\omega ^{2}\ n_{1}^{\ }h_{4} & \omega ^{2}\ n_{2}^{\ }h_{4} & 0 & h_{4}%
\end{array}%
\right| ,
\end{equation*}%
 where, in this work, $y^{a}=(y^{3}=t,y^{4}=y)$ and spacetime
sig\-na\-tu\-re is chosen $(+,+,-,+).$ The coefficients $%
g_{k}(x^{i}),h_{a}(x^{i},t),$ $w_{k}(x^{i},t),n_{k}(x^{i},t),q(x^{i},t)$ and
$\omega (x^{i},t,y)$ can be defined in explicit form by integrating and/or
differentiating some generating functions. For instance, the class of
certain solutions (\ref{gsol}) is with functional dependence on a generating
function $\phi (x^{k},t),$ when $h_{a}=h_{a}[x^{i},\phi ]$ and (following
from relations for v--metric) $w_{i}=w_{i}[x^{i},t,\phi ]$ and $%
n_{i}=n_{i}[x^{i},t,\phi ].$ The type of chosen generating and integration
functions (smooth class, sure or stochastic character), prescribed
symmetries and topology of interactions, boundary/limit conditions etc
distinguish in explicit form a geometric/physically important class of
solutions.
\end{itemize}

\item \textbf{Elaborate a nonholonomic stochastic calculus and related
general relativistic diffusion theory on} $\mathbf{V}$

\begin{itemize}
\item The theory of stochastic processes and diffusion on curved manifolds,
Riemann and Riemann--Cartan spaces \footnote{%
the last ones are with nontrivial torsion; torsion can be also induced
nonholonomically on Minkowski and Riemann spaces but there are canonical
anholonomic transforms to the Levi--Civita connection with zero torsion},
and bundle spaces is a well developed direction in modern mathematics.%
\footnote{%
During last two decades the approach was extended for (in general,
suppersymmetric) Lagrange--Finsler spaces, and higher order generalizations,
and on (pseudo) Riemannian/Lorentz manifolds with various applications of
relativistic diffusions in modern cosmology and astrophysics.}

\item In this work, we have shown how the mathematical formalism for the
relativistic stochastic theory and diffusion can be adapted for nonholonomic
manifolds with nontrivial N--connection structure. In brief, the
constructions are those for (pseudo) Riemannian manifolds but with
corresponding generalizations with respect necessary classes of
orthonormalized N--adapted frames, modified linear connections and
Laplace--Beltramy operators.

\item The geometric data $(\mathbf{g,N,}\widehat{\mathbf{D}})$ and/or sure
general solutions are considered as a fixed nonholonomic background for
''non--integrab\-le'' rolling of stochastic processes in tangent bundles
with h-- and v--splitting. So, at this stage the generating/integration
functions for the solutions of Einstein equations are taken to be sure
functions even fluctuations to nonholonomic Einstein--Langeven systems,
induced by quasi--classical quantum fluctuations of matter, can be
considered in a self--consistent form as in so--called stochastic gravity %
\cite{hu}.
\end{itemize}

\item \textbf{Consider random generating functions and sources for formal
solutions of Einstein equations}

\begin{itemize}
\item The key idea of this work is to consider (pseudo) Riemannian metrics
with coefficients generalized in a form to include random variables induced
by generating functions (for some particular cases, we can consider directly
some N--adapted coefficients of v--metric, $h_{ab},$ and N--connection, $%
N_{i}^{a})$ defined, for instance, by some gravitational diffusion processes.

\item Transitions from a ''sure'' nonholonomic (pseu\-do) Riemannian
configuration $(\ ^{\circ }\mathbf{g,\ ^{\circ }N,}\ ^{\circ }\widehat{%
\mathbf{D}}_{\mid Z=0}\rightarrow \ ^{\circ }\nabla)$ to a stochastic
nonholonomic Einstein spacetime $(\ ^{\eta }\mathbf{g,}\ ^{\eta }\mathbf{N,}%
\ ^{\eta }\widehat{\mathbf{D}}_{\mid Z=0}\rightarrow \ ^{\eta }\nabla)$ are
modeled by metrics of type (\ref{gsol}), when $g_{ij}=diag[g_{i}=\eta _{i}\
^{\circ }g_{i}]$ and $h_{ab}=diag[h_{a}=\eta _{a}\ ^{\circ }h_{a}]$ and $%
N_{k}^{3}=w_{i}=\eta _{i}^{3}\ ^{\circ }w_{i}$ and $N_{k}^{4}=n_{i}=\eta
_{i}^{4}\ ^{\circ }n_{i}$ are constructed for some sure and/or stochastic
gravitational $\eta $--polarizations. The polarizations are defined in such
a way that $\ ^{\eta }\mathbf{g}$ generates a class of exact solutions of
the Einstein equations with $\eta _{\alpha }$ and $\eta _{i}^{a}$ being
related to some stochastic/diffusion processes (as solutions of some
generalized and/or nonholonomically constrained Kolmogorov/ Fokker--Planck
equations) rolled on a ''sure'' background $(\ ^{\circ }\mathbf{g,\ ^{\circ
}N,}\ ^{\circ }\widehat{\mathbf{D}}).$ We have to change the usual ''sure''
integrations into stochastic Stratonovich ones (which is the most convenient
for curved spaces), with possible re--definition of results for It\^{o}
stochastic integrals. For real physical applications, we have to introduce
orthonormalized N--adapted frames related to some certain degrees defined by
$\ ^{\circ }\mathbf{g}$ and sure deformations to $\ ^{\eta }\mathbf{g}$ but
consider that additional deformations are driven by some Wiener processes $%
\delta \mathcal{W}^{\widehat{a}}$ (for physicists, all necessary concepts on
stochastic calculus and diffusion are summarized in Ref. \cite{lemons}).

\item In explicit form, the nonholonomic transitions between sure and
stochastic gravitational configurations are modelled by changing of a
''sure'' generating function into a random one, $\phi (x^{k},t)\rightarrow
\underline{\phi }(x^{k},t)=\phi (x^{k},t)+\varpi \widetilde{\phi }(x^{k},t),$
see (\ref{gsol1}). Such mutual \ evolutions of deterministic and random
phases are possible even for ''sure'' boundary/initial conditions, in
classical gravity because of generic nonlinear character of Einstein
equations. The stochastic behavior may arise for very small values of
parameter $\varpi $ and vanish if $\varpi =0.$ Geometrically, we can impose
such nonholonomic splitting into conventional h-- and v--subsystems with
distinguished sure and stochastic variables, which allows us to split the
equations and generate exact solutions.
\end{itemize}
\end{enumerate}

We emphasize here that it would not be possible to elaborate such, in
general, non--perturbative methods of constructing exact solutions in
gravity with ''sure'' and/or ''stochastic'' variables, and their mixture, if
we do not apply corresponding methods from the geometry of nonholonomic
manifolds/bundles and N--connection formalism (originally proposed in
Finsler and Lagrange geometry and further re--considered in Einstein gravity
and modifications).

Let us compare our methods to an alternative approach in stochastic gravity %
\cite{hu}. For various applications in gravity and \ astrophysics, there are
considered semiclassical generalizations of Einstein equation with sources $%
\langle T_{ab}[g]\rangle _{ren}^{\prime }$ computed for suitably
renormalized expectation value of the stress tensor operator. Such a
renormalized value is computed with the scalar field operator satisfying the
Klein--Gordon equation on the perturbed metric $g_{ab}+h_{ab}$ and
stochastic source $\xi _{ab}.$ We get the so--called Einstein--Langevin
equation,
\begin{equation*}
\ ^{1}G_{ab}[g+h]=\varkappa \langle \ ^{1}\widehat{T}_{ab}[g+h]\rangle
_{ren}+\varkappa \xi _{ab}[g],
\end{equation*}%
where the left superindex means that only terms linear in the metric
perturbations are kept (the stochastic source $\xi _{ab}$ is regarded to be
of the same order as $h_{ab}$). The Gaussian stochastic source $\xi _{ab}$
is completely determined by the following correlation functions:%
\begin{equation*}
\langle \xi _{ab}[g;x)\rangle _{\xi }=0,\ \ \langle \xi _{ab}[g;x)\xi
_{ab}[g;y)\rangle _{\xi }=N_{abcd}(x,y)=\frac{1}{2}\langle \left\{ \widehat{t%
}_{ab}[g;x),\widehat{t}_{cd}[g;y)\right\} \rangle ,
\end{equation*}%
where $\langle \ldots \rangle _{\xi }$ denotes the expectation value with
respect to the stochastic classical (i.e. non--quantum) source $\xi
_{ab}[g]. $ The operator $\widehat{t}_{ab}[g]$ is \ defined as $\widehat{t}%
_{ab}[g]:=\widehat{T}_{ab}[g]-\langle \widehat{T}_{ab}[g]\rangle $ and the
bitensor $N_{abcd}(x,y),$ which determines the correlation function of the
stochastic source, is computed using the scalar field operator satisfying
the Klein--Gordon equation for the background metric $g_{ab}.$ The bitensor $%
N_{abcd}(x,y)$ is called the noise kernel, it describes the quantum
fluctuations of the stress tensor operator and is positive--semidefinite.

Comparing our approach and the stochastic gravity scheme we conclude that we
can include that scheme into the anholonomic deformation method considering
a subclass of perturbative ''fluctuations'' of metrics defined by ''noise''
in energy--momentum tensors. In general, following stochastic nonholonomic
geometric constructions we can model generic nonlinear stochastic processes
both in vacuum and non--vacuum gravity.

\vskip3pt

{\small
\textbf{Acknowledgement: } I'm grateful to F. Mainardi, N. Mavromatos  and P.
Stavrinos for important  discussions, support and collaboration. The research in this paper is partially supported by the Program IDEI, PN-II-ID-PCE-2011-3-0256.}

\end{document}